\documentclass[sn-mathphys]{sn-jnl}

\usepackage{graphicx}%
\usepackage{multirow}%
\usepackage{amsmath,amssymb,amsfonts}%
\usepackage{amsthm}%
\usepackage{mathrsfs}%
\usepackage[title]{appendix}%
\usepackage{xcolor}%
\usepackage{textcomp}%
\usepackage{manyfoot}%
\usepackage{booktabs}%
\usepackage{algorithm}%
\usepackage{algorithmicx}%
\usepackage{algpseudocode}%
\usepackage{listings}%

\newcommand{\R}{\mathbb{R}}
\newcommand{\X}{\mathcal{X}}
\newcommand{\E}{\mathbb{E}}

\newcommand{\V}{\mathbb{V}}

\newcommand{\argmin}{\operatornamewithlimits{argmin}}
\DeclareMathOperator*{\argmax}{\arg\!\max}

\begin{document}

\title[Understanding an Acquisition Function Family for Bayesian
Optimization]{Understanding an Acquisition Function Family for Bayesian
Optimization}


\author[1]{\fnm{Jiajie} \sur{Kong}}\email{jkong7@ucsc.edu}
\equalcont{These authors contributed equally to this work.}

\author*[2]{\fnm{Tony} \sur{Pourmohamad}}\email{pourmoht@gene.com}
\equalcont{These authors contributed equally to this work.}

\author[1]{\fnm{Herbert K. H.} \sur{Lee}}\email{herbie@ucsc.edu}
\equalcont{These authors contributed equally to this work.}

\affil*[1]{\orgdiv{Department of Statistics}, \orgname{University of California, Santa Cruz}, \orgaddress{\street{1156 High Street}, \city{Santa Cruz}, \postcode{95064}, \state{CA}, \country{USA}}}

\affil[2]{\orgdiv{Data and Statistical Sciences}, \orgname{Genentech}, \orgaddress{\street{1 DNA Way}, \city{South San Francisco}, \postcode{94080}, \state{CA}, \country{USA}}}


\abstract{Bayesian optimization (BO) developed as an approach for the efficient optimization of expensive black-box functions without gradient information. A typical BO paper introduces a new approach and compares it to some alternatives on simulated and possibly real examples to show its efficacy. Yet on a different example, this new algorithm might not be as effective as the alternatives. This paper looks at a broader family of approaches to explain the strengths and weaknesses of algorithms in the family, with guidance on what choices might work best on different classes of problems.}

\keywords{Black-box function , expected improvement, improvement function, Gaussian processes}



\maketitle

\section{Introduction}
\label{sec:intro}

Expensive black-box functions arise in many scientific disciplines where computer models are needed to model complex physical systems \citep{gramacy:2020,pourmohamad:book:2021}. These computer models (or black-box computer codes) are typically deployed when direct experimentation of the physical system under study is prohibitive. For example, geological phenomena, such as earthquakes and volcanic eruptions, are not reproducible physical experiments, and so computer models based on seismology and volcanology are sometimes used to study these events. Typically, the black-box functions describing the complex system under study are highly complex, multi-modal, and difficult to understand, which makes optimizing these black-box functions a challenging problem. The optimization problem becomes even more difficult when the black-box functions are computationally expensive to evaluate and no gradient information is available. Given the computational expense of evaluating these types of black-box functions, there is a clear need for efficient sequential optimization algorithms that do not require many function evaluations. 

In the context of optimizing expensive black-box functions, a popular solution for this type of problem is to use Bayesian optimization (BO) \citep{mockus:1978}. BO is an efficient sequential design strategy for optimizing expensive black-box functions, in few steps, that does not require gradient information \citep{brochu:2010}. More precisely, BO is well suited for solving optimization problems of the following form
\begin{align}\label{BO:problem}
    x^*\in\argmin_{x\in\X} f(x)
\end{align}
where $\X\subset \R^d$ is a known, bounded region such that $f:\X\rightarrow\R$ denotes a scalar-valued objective function. Here, we regard $f(x)$ as the output of evaluating the objective function at input $x$. Furthermore, we treat $f(x)$ as a black-box function that only returns function evaluations of the objective function $f$ and does not provide any gradient information about it, i.e., we focus on the case of derivative free optimization \citep{conn:scheinberg:vicente:2009}. BO proceeds in solving (\ref{BO:problem}) by iteratively developing a ``cheap-to-compute" model, or {\em surrogate model} \citep{gramacy:2020}, of the objective function $f$, and at each step of this iterative process, using predictions from the surrogate model to maximize an acquisition (or utility) function, $a(x)$, that measures how promising each location, $x\in\X$, in the input space is if it were to be the next chosen point to evaluate. 

Clearly, the success of the BO algorithm is heavily tied to the efficiency of the acquisition function for guiding the search \citep[to name a few]{scho:welc:jone:1998,tadd:lee:gray:grif:2009, srinivas:2010, snoek:2012, henning:shuler:2012, hernandez:lobato:2014}. A good acquisition function
should accurately reflect our beliefs about which is the best next input to evaluate, while also striking a balance between exploration (global search) and exploitation (local search). With this in mind, one of the most widely used acquisition functions in BO was developed by \cite{jones:schonlau:welch:1998}, namely the expected improvement (EI) acquisition function (Section \ref{sec:EI}). As the name suggests, a new candidate input, $x$, is chosen such that it maximizes the expected improvement (i.e., reduction) in the solution to the optimization problem in (\ref{BO:problem}) over other possible candidate inputs. Here, expected is synonymous with average and so the EI acquisition function can be viewed as a point estimate of the average improvement. A natural extension would then be to think about quantifying the uncertainty (say with a confidence or credible interval, for example) for the point estimate, yet, little work in this direction has been done \citep{noe:2019, marisu:2021}. To this end, this paper proposes a new BO acquisition function that naturally, and efficiently, incorporates the associated uncertainty for the point estimate of the EI acquisition function. The development of this new acquisition function will also lead to a general framework for constructing an acquisition function family for dealing with uncertainty in the EI acquisition function. 

We emphasize that the primary goal of this paper is to provide intuition for parameters in a family of acquisition functions, to better explain how different acquisition functions work and on what types of problems each one will work best. This is not intended to be another paper that introduces a new acquisition function and attempts to show that it is better than existing functions. We are interested in understanding existing and new functions. As no acquisition function is optimal on every possible problem, it is helpful to know which acquisition function is best suited for different types of problems.

The remainder of this article is organized as follows. In Section \ref{sec:cbo}, we introduce the two integral components for understanding the inner working of a BO algorithm. Section \ref{sec:quantifying} outlines the concept of expected improvement, how and why we might quantify the uncertainty in our improvement, and establishes an acquisition function family based on improvement functions. Section \ref{sec:examples} details the strengths and weaknesses of different algorithm settings in the acquisition function family based on simulated examples. Lastly, Section \ref{sec:disc} finishes with some discussion.

\section{Components of Bayesian Optimization}
\label{sec:cbo}
Section \ref{sec:cbo} introduces the two main components that are essential for conducting Bayesian optimization.
\subsection{Acquisition Functions}
The performance of any BO algorithm is inherently tied to its acquisition function. An acquisition function, $a_n(x)$, encodes a measure of belief of how promising an input $x$ is at minimizing the objective function $f(x)$ in (\ref{BO:problem}). At the $n^{th}$ iteration of the BO algorithm, the best next input $x_n$ to evaluate is chosen such that
\begin{align}\label{acq}
x_n = \argmax_{x\in\X} a_{n-1}(x).
\end{align}
Here, the strategy in (\ref{acq}) is to choose the input that maximizes the acquisition function since the maximizing the acquisition function is akin to maximizing our beliefs about where the best next input to evaluate is for minimizing the problem in (\ref{BO:problem}). 

Obviously, different choices (i.e., functional forms) of acquisition function will inherently lead to different beliefs in which is the best next input to evaluate, and many are suggested in the literature \citep[for example,][]{scho:welc:jone:1998, jones:schonlau:welch:1998,frazier:2008,srinivas:2010,henning:shuler:2012,kandasamy:2018}. However, all useful acquisition functions have one feature in common which is that they make use of an exploitation-exploration trade-off. The exploitation-exploration trade-off says that a good acquisition function should trade-off between searching the input space globally (exploration) and searching the input space locally (exploitation). Too much exploration and the BO algorithm will likely not converge to a solution for (\ref{BO:problem}), and too much exploitation leads to BO algorithms that tend to get stuck in local modes (i.e., local minima) of the input space and thus never finding the global solution to (\ref{BO:problem}). 

A popular choice of acquisition function that directly enforces the exploitation-exploration trade-off is the EI acquisition function \citep{jones:schonlau:welch:1998}. This acquisition function provides a basis for exploration of a family of other acquisition functions, and it also serves as a benchmark for the effectiveness of other functions. The finer details of the EI acquisition function are discussed in Section \ref{sec:EI}.

\subsection{Gaussian Process Surrogate Modeling}
Once an acquisition function for BO has been chosen, the next step is to develop a strategy for maximizing the acquisition function.
For the optimization problem in (\ref{acq}), the BO algorithm essentially embeds an optimization problem inside of an already challenging and computationally expensive optimization problem in (\ref{BO:problem}). Given these computational challenges associated with (\ref{BO:problem}), it is necessary that the optimization problem in (\ref{acq}) be a much easier and faster problem to solve. With the expensive nature of evaluating inputs $x$, the BO algorithm relies on developing a surrogate model \citep{gramacy:2020, pourmohamad:book:2021} that relates the inputs $x$ to the outputs $f(x)$ and can be used to make predictions of the outputs, say $f(x^*)$, at untried inputs, $x^*$. The typical choice of surrogate model for BO has been the Gaussian process (GP) \citep{sant:will:notz:2003}. GPs are typically viewed as a highly flexible nonparametric regression model which, when acting as a surrogate model, are much faster/cheaper to predict untried inputs when compared to evaluating the actual objective function $f$. Moreover, GP surrogate models allow for uncertainty quantification in the prediction of the objective function at untried inputs which tends to be a critical component of most acquisition functions.

Fundamentally, GPs are distributions over functions such that the joint distribution at any finite set of points is a multivariate Gaussian distribution, and are defined by a mean function and a covariance function. Let $\{x^{(i)},y^{(i)}\}^{n}_{i=1}$ denote the input-output pairs of data after $n$ input evaluations of the objective function $f$. The GP, $Y(x)$, serves as the surrogate model for the data  $\{x^{(i)},y^{(i)}\}^{n}_{i=1}$ and its predictive equations can be obtained as a straightforward consequence of conditioning for multivariate normal joint distributions, that is, the predictive distribution $Y(x)|\{x^{(i)},y^{(i)}\}^{n}_{i=1}$ at a new input $x$ follows another Gaussian process $Y(x)|\{x^{(i)},y^{(i)}\}^{n}_{i=1}\sim N(\mu(x),\sigma^2(x))$. The choice of surrogate model that we use for the remainder of this paper is the GP.

\section{Quantifying the Improvement}
\label{sec:quantifying}

\subsection{Expected Improvement}\label{sec:EI}
Originally introduced by \cite{jones:schonlau:welch:1998} in the computer modeling literature, the improvement function, $I(x)=\max_x\{0,f_{\min}^n-Y(x)\}$, measures the amount of improvement of an untried input, $x$, over the current observed minimum value $f_{\min}^n = \min\{f(x_1),...,f(x_n)\}$ after $n$ runs of the computer model. Since the untried input $x$ has not yet been observed, both $Y(x)$ and $I(x)$ are unknown and can be regarded as random variables. Here, the usual approach is to model $Y(x)$, conditional on the observed inputs $x_1,...,x_n$, using a Gaussian process surrogate model. Under this assumption, one can calculate the expectation of the improvement function, or rather the expected improvement acquisition function, i.e., 
\begin{align}\label{EI}
\text{EI}(x)=\E(I(x)) = (f^n_{\min} - \mu_n(x))\Phi\left(\frac{f_{\min}^n-\mu_n(x)}{\sigma_n(x)}\right) + \sigma_n(x)\phi\left(\frac{f_{\min}^n-\mu_n(x)}{\sigma_n(x)}\right)
\end{align}
where $\mu_n(x)$ and $\sigma_n(x)$ are the mean and standard deviation of the predictive distribution of $Y(x)$, and $\Phi(\cdot)$ and $\phi(\cdot)$ are the standard normal cdf and pdf, respectively.

Conceptually, the EI acquisition function makes a lot of sense. We should intuitively favor trying new candidate inputs $x$ where we expect, on average, for the improvement to be high over other solutions to (\ref{BO:problem}). Moreover, the form of the EI acquisition function in (\ref{EI}) provides a combined measure of how promising a candidate input is by trading off between local search ($\mu(x)$ under $f_{\min}$) and global search ($\sigma(x)$).

\subsection{Variance of the Improvement}\label{sec:varimprov}
The EI acquisition function grew organically out of the intuitive notion that candidate inputs should be chosen based on where we should expect, on average, for the improvement to be high. However, when recalling the definition of EI in (\ref{EI}), the EI is the expectation of the improvement function, or rather a point estimate of the random function $I(x)$ and so there is a quantifiable amount of uncertainty associated with our point estimate as well. In order to understand the variability (or uncertainty) associated with the EI acquisition function, one needs to calculate the variance of the improvement function $I(x)$. Fortunately, the variance of the improvement function, under the assumption of a GP surrogate model, has the following closed form expression: 
\begin{align}\label{VI}
    \text{VI}(x) = \V\text{ar}(I(x)) &= \sigma_n^2(x)\Bigg[\left(\left(\frac{f_{\min}^n-\mu_n(x)}{\sigma_n(x)}\right)^2 + 1\right)\Phi\left(\frac{f_{\min}^n-\mu_n(x)}{\sigma_n(x)}\right) +\\ 
    & \left(\frac{f_{\min}^n-\mu_n(x)}{\sigma_n(x)}\right)\phi\left(\frac{f_{\min}^n-\mu_n(x)}{\sigma_n(x)}\right)\Bigg] - (\text{EI}(x))^2. \nonumber
\end{align}
The details of the derivation of $\text{VI}(x)$ can be found in \cite{scho:welc:jone:1998}. Interestingly, most works in the BO literature have focused mainly on expected improvement as is, with no regards to direct uncertainty quantification in the improvement function. \cite{scho:welc:jone:1998} was the first to calculate the variance of the improvement function, but quantifying the variability in the improvement was not the main goal of the paper, but rather a result that fell out of their methodology of calculating the power expected improvement (PEI), i.e., $\E(I^g(x))$ for $g > 0$. The case of $g=2$ leads to the derivation of the variance of the improvement function, $\text{VI}(x)$, since $\text{VI}(x) = \E(I^2(x)) - [\E(I(x))]^2$. 

It was not until recently though that any BO algorithms made any attempt to consider incorporating the variance of the improvement function into the acquisition function. In particular, \cite{noe:2019} and \cite{marisu:2021} take two different approaches to incorporating the uncertainty in the improvement function into their respective acquisition functions. \cite{noe:2019} introduced the concept of the scaled expected improvement (SEI) acquisition function as the following:
\begin{align}\label{SEI}
    \text{SEI}(x) = \frac{\text{EI}(x)}{\sqrt{\text{VI}(x)}}.
\end{align}
The acquisition function in (\ref{SEI}) scales the expected improvement by the reciprocal of the standard deviation of the improvement function, and by doing so, attempts to create an acquisition function that corresponds to selecting inputs where the improvement is expected to be high with high certainty. However, it should be noted that the SEI acquisition function may lead to a BO algorithm that overly favors local search since SEI will be maximized at or near points where the variance of the improvement function is close to 0, which commonly occurs at or around inputs that have already been previously evaluated (i.e., areas of high exploitation rather than exploration). On the other hand, \cite{marisu:2021} introduced an acquisition function (which we refer to as VEI) as a linear combination of the expectation and variance of the improvement function, i.e., 
\begin{align}\label{VEI}
    \text{VEI}(x) = \text{EI}(x) - \frac{\xi}{2}\text{VI}(x).
\end{align}
Here, $\xi>0$ can be thought of as a tuning parameter that controls the amount of uncertainty in which to penalize the expected improvement by as well rates of convergence of the BO algorithm (similar to the tuning parameter found in the upper confidence bound acquisition function of \cite{srinivas:2010}). Likewise, the choice of $\xi$ may change during the course of running the BO algorithm and may as, say, a function of the total number of inputs evaluated. However, based upon empirical evidence, \cite{marisu:2021} recommend setting $\xi=1$ given their experience working with the VEI acquisition function. Note that there is no theoretical guarantee that VEI has to be greater than 0. In fact, it is easy to see that when $(\xi/2)\text{VI}(x)>\text{EI}(x)$ for all $x$, that VEI will either be a negative number, or VEI will be 0 at inputs that have already been evaluated. However, under the very real scenario that $(\xi/2)\text{VI}(x)>\text{EI}(x)$ for all $x$, this is problematic since maximizing the VEI acquisition function under this scenario will lead to choosing inputs that are close in proximity to previously evaluated inputs which can lead to a local search algorithm that will likely get stuck in local minima of the objective function. With these problems in mind, the next section introduces a new acquisition function intended to address these issues.

\subsection{Accounting for Uncertainty}
Although \cite{noe:2019} and \cite{marisu:2021} describe separately different ways to account for the uncertainty in the improvement function when using expected improvement, as previously discussed, there are some clear deficiencies in both their SEI and VEI acquisition functions. More specifically, there are cases where SEI and VEI break down due to small or large variances in the improvement function, respectively. In order to overcome some of these issues, we propose a new acquisition function which accounts for the uncertainty in the expected improvement without encountering these local search issues. Here we define an acquisition function based on uncertainty in the expected improvement (referred to as UEI) as follows: 
\begin{align}\label{UEI}
\text{UEI}(x) = \text{EI}(x) + \gamma \sqrt{\text{VI}(x)},
\end{align}
where $\gamma > 0$. Given that we desire uncertainty quantification around the expected improvement, the form of the UEI acquisition in (\ref{UEI}) feels like a natural choice given that it resembles what would be a credible (or confidence) interval for the point estimate of the expected improvement. That is to say, for appropriate choices of $\gamma$, the UEI acquisition function can be viewed as maximizing the upper quantiles of a $(1-\alpha)$\% credible (or confidence) interval, $\alpha\in(0,1)$, for the EI. 

The individual components of the UEI acquisition function may not look drastically different than that of the SEI and VEI acquisition functions, however, the UEI acquisition function incorporates the variance in the improvement function in a very different manner. Both the SEI and VEI acquisition functions penalize the EI for having high variability in the improvement function, which sounds natural if we are interested in choosing new inputs based on high EI and high certainty in the EI. However, this penalization ultimately leads to BO algorithms which may display a higher degree of local, rather than global, search when optimizing the objective function. On the other hand, the UEI acquisition function favors rewarding variability in the improvement function which is beneficial for two reasons. First, since $\gamma \sqrt{\text{VI}(x)} \geq0$ for all $x$, the UEI acquisition will always be a non-negative value and so it will not suffer from the same local search issues that plague VEI. Likewise, when the variability in the improvement function is large, UEI will tend to favor global search, however, as the variability in the improvement goes to 0, the UEI will also not suffer from the local search issues encountered by SEI since the UEI acquisition function will converge to the original EI acquisition function in (\ref{EI}) as $\text{VI}(x)$ goes to 0. Secondly, the form of the UEI acquisition function suggests treating the search for the best next input to evaluate less pessimistically that SEI and VEI. By this we mean that the UEI acquisition function suggests picking the best next input as the one that will give the highest potential expected improvement as measured by the upper credible interval of the EI, as opposed to focusing in on areas of high expected improvement with high certainty. The reward for embracing uncertainty in this way is that UEI will function as an acquisition function that can still efficiently balance the exploration-exploitation trade-off. 

As shown in \cite{noe:2019} and \cite{marisu:2021}, the SEI and VEI acquisition functions are not without their merits. In fact, we believe that there does not exist a single best acquisition function for every scenario, but that the SEI and VEI acquisition functions perform better under certain scenarios.  With this in mind, we envision that there is a general acquisition function family that encapsulates the class of expected improvement based acquisition functions, and that the general form for the family of acquisition functions may provide insights into when one EI-based acquisition function is preferable to another. And so, we define an acquisition function family using the following acquisition function: 
\begin{align}\label{generalform}
    a(x) = \frac{\E(I^w(x))}{[\V\text{ar}(I(x))]^u} + \beta [\V\text{ar}(I(x))]^v
\end{align}
with $\beta\in\mathbb{R}$, and $ u,v,w \geq 0$. The acquisition function in (\ref{generalform}) incorporates components of both the expectation and variance of the improvement function, while the parameters $u, v,$ and $w$ govern how much of each component to use, and if the acquisition function should be a linear combination or scaling of the uncertainty in the improvement function, or both. It is obvious to see that for certain choices of $u$, $v$, $w$, and $\beta$, the acquisition function in (\ref{generalform}) will recover the EI, PEI, SEI, VEI, and UEI acquisition functions exactly (Table \ref{tab:summary}). For PEI, any positive $w$ could be used, and $w=2$ is the most commonly used value, so we use that for the rest of this paper.
\begin{table}[htb]
\begin{tabular}{l c c c c c}\hline
Source & Acquisition Function & $u$ & $v$ & $w$ & $\beta$ \\
\hline
\cite{jones:schonlau:welch:1998} & EI & 0 & -- & 1 & 0 \\
\cite{scho:welc:jone:1998} & PEI &0 & -- & 2 & 0 \\
\cite{noe:2019} & SEI & 1/2 & -- & 1 & 0 \\
\cite{marisu:2021} & VEI & 0 & 1 & 1 & $<0$ \\
-- & UEI & 0 & 1/2 & 1 & $>0$ \\
\hline
\end{tabular}
\caption{The parameter settings of $u,v,w$ and $\beta$ for recovering the different acquisition functions from the general form for the acquisition function family.}
\label{tab:summary}
\end{table}

We discuss the roles that each value $u$, $v$, $w$, and $\beta$ play in the general form for the acquisition function family in the next section. 

\section{Illustrative Examples}
\label{sec:examples}
To demonstrate the performance, strengths, and weaknesses of the different acquisition functions, we solve several well-known optimization problems in Section \ref{opt:problems} using the EI, PEI, SEI, VEI, and UEI acquisition functions, and compare and contrast their respective results. Section \ref{sec:affp} explores the values of the parameters in the functional family. For VEI, $\beta$ is a free parameter, and we focus primarily on the recommended value of $\beta=-1/2$. For UEI, we have tried a variety of values and found that $\beta = 2$ generally works well.

\subsection{Optimization Test Problems}\label{opt:problems}
Given the popularity of the EI acquisition function, its performance on optimization test functions is well-known, yet the effect on this performance given the addition of quantifying the uncertainty in the improvement function through its variance is much less known.  In this section, we seek to minimize six different optimization test functions taken from the optimization community \citep{simulationlib} via the five different acquisition functions. Our choice of optimization test functions were based on choosing optimization problems that contained objective functions with either many local minima, solutions that lied along the boundary of the input space, valley shapes, steep ridges or drops, or any combination of these qualities, in order to induce different levels of difficulty for each of the acquisition functions. Further characteristics of the optimization test functions can be found in Table \ref{tab:problems}, and the exact form of the equations for the test functions can be found in \ref{append}.  

\begin{table}[htb]
\begin{tabular}{l c c c c}\hline
Test Function & Abbreviation & Number of & Number of & Global Solution\\
& & Dimensions & Local Minima & $f(x)$\\
\hline
Gramacy and Lee & GRL & 1 & 10 & -0.869 \\
Rosenbrock & ROS & 2 & 1 & 0\\
Modified Townsend & MOT &  2 & 6 & -2.969 \\
Ackley & ACY & 2 & 25 & 0 \\
Rastrigin & RAS & 2 & 25 & 0 \\
Hartman & HTN & 6 & 6 & -3.322\\
\hline
\end{tabular}
\caption{The different optimization test functions used to evaluate the performance of the different acquisition functions.}
\label{tab:problems}
\end{table}

\begin{figure}[htbp]
    \centering
    \begin{tabular}{cc}
    \includegraphics[scale = .13]{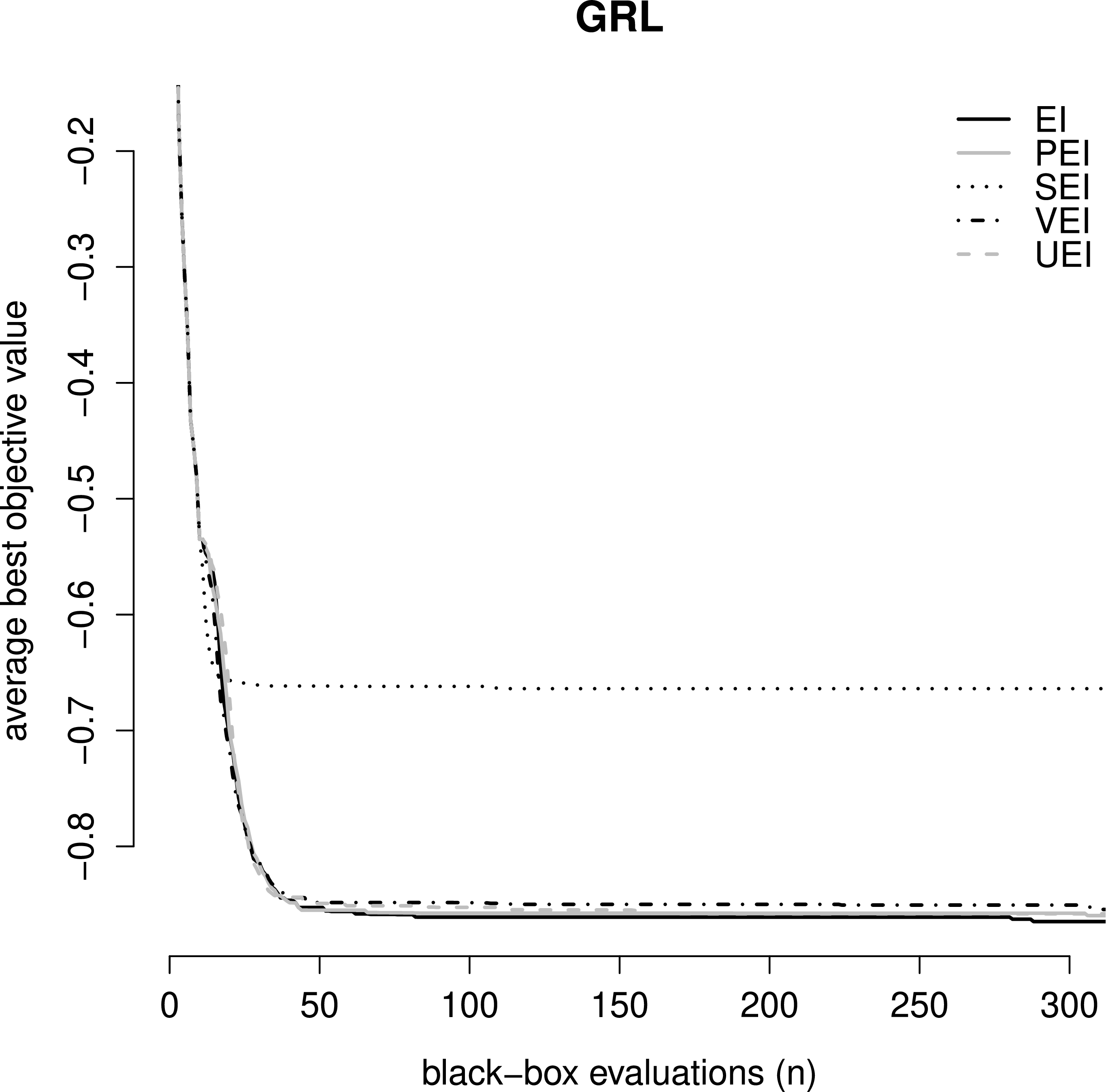}
    \includegraphics[scale = .13]{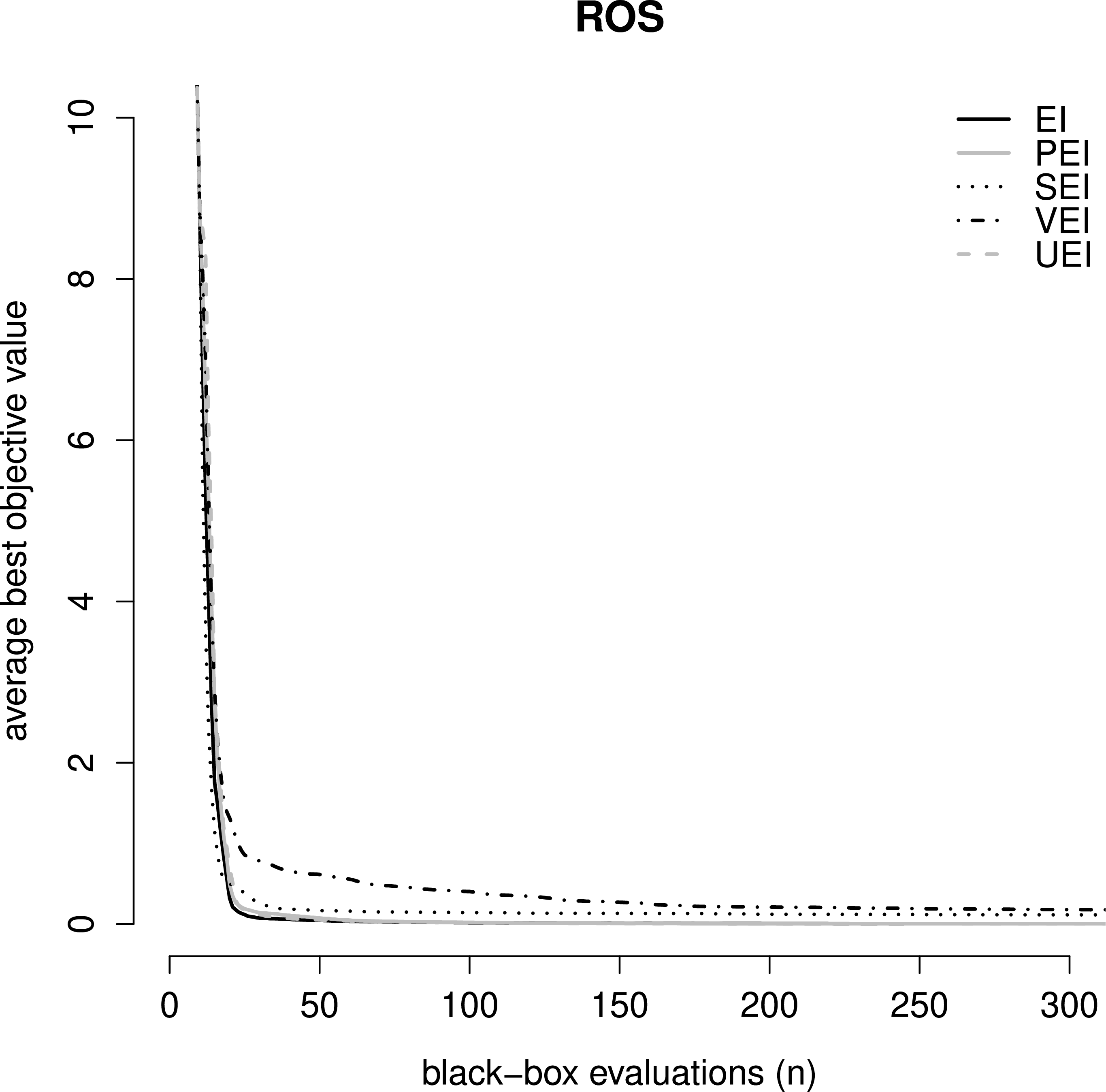}\\
    \includegraphics[scale = .13]{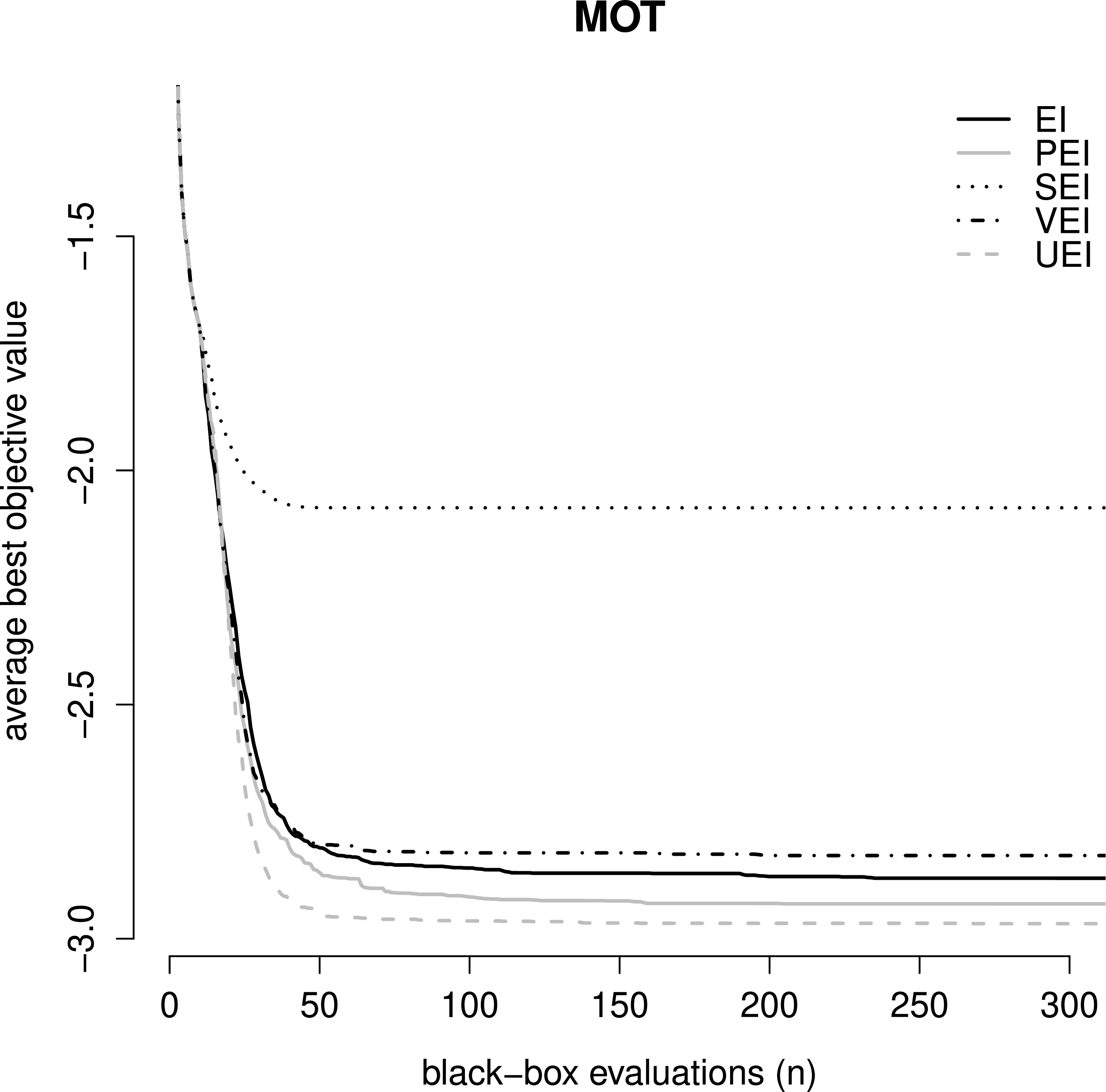}
    \includegraphics[scale = .13]{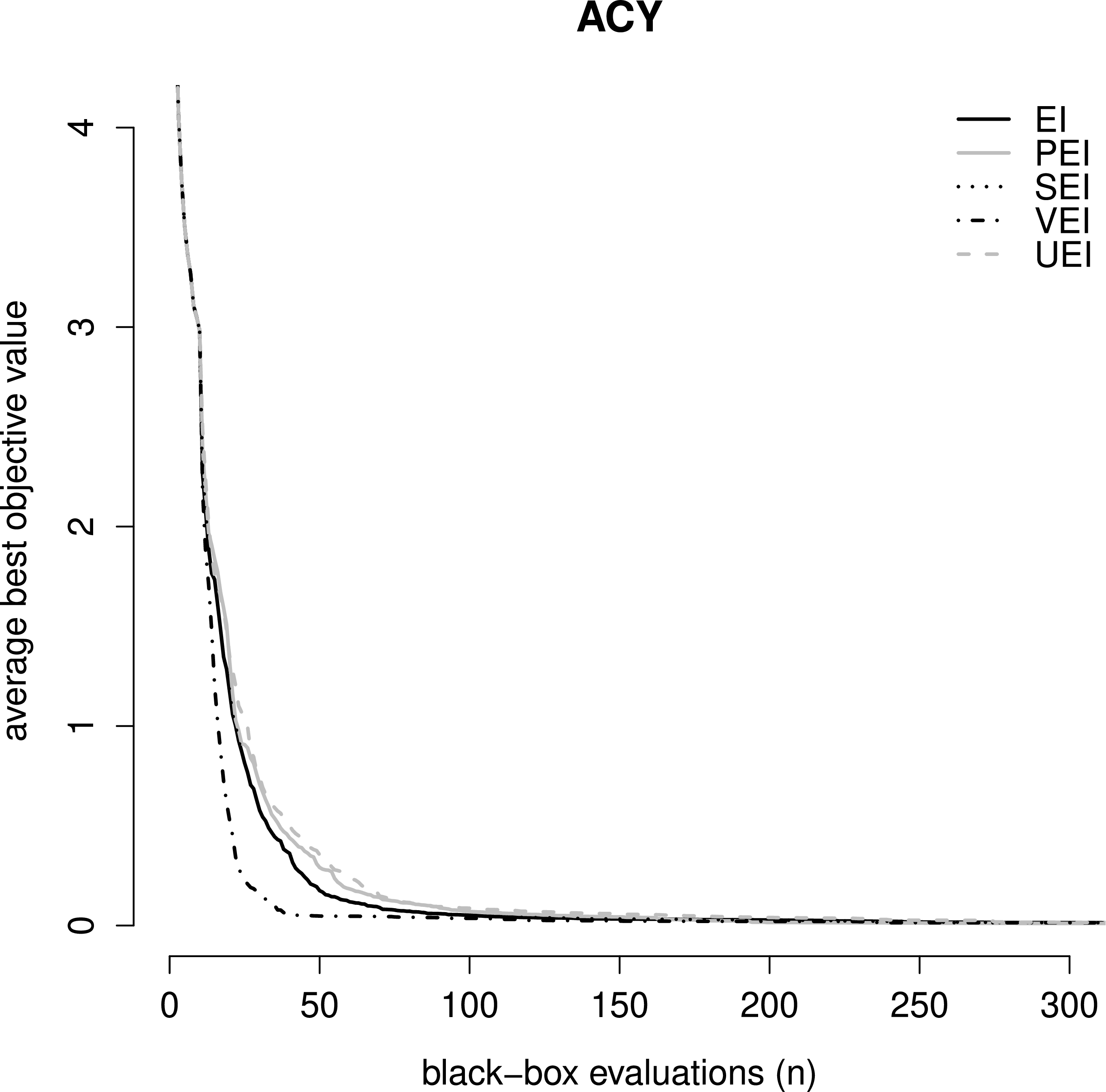}\\
    \includegraphics[scale = .13]{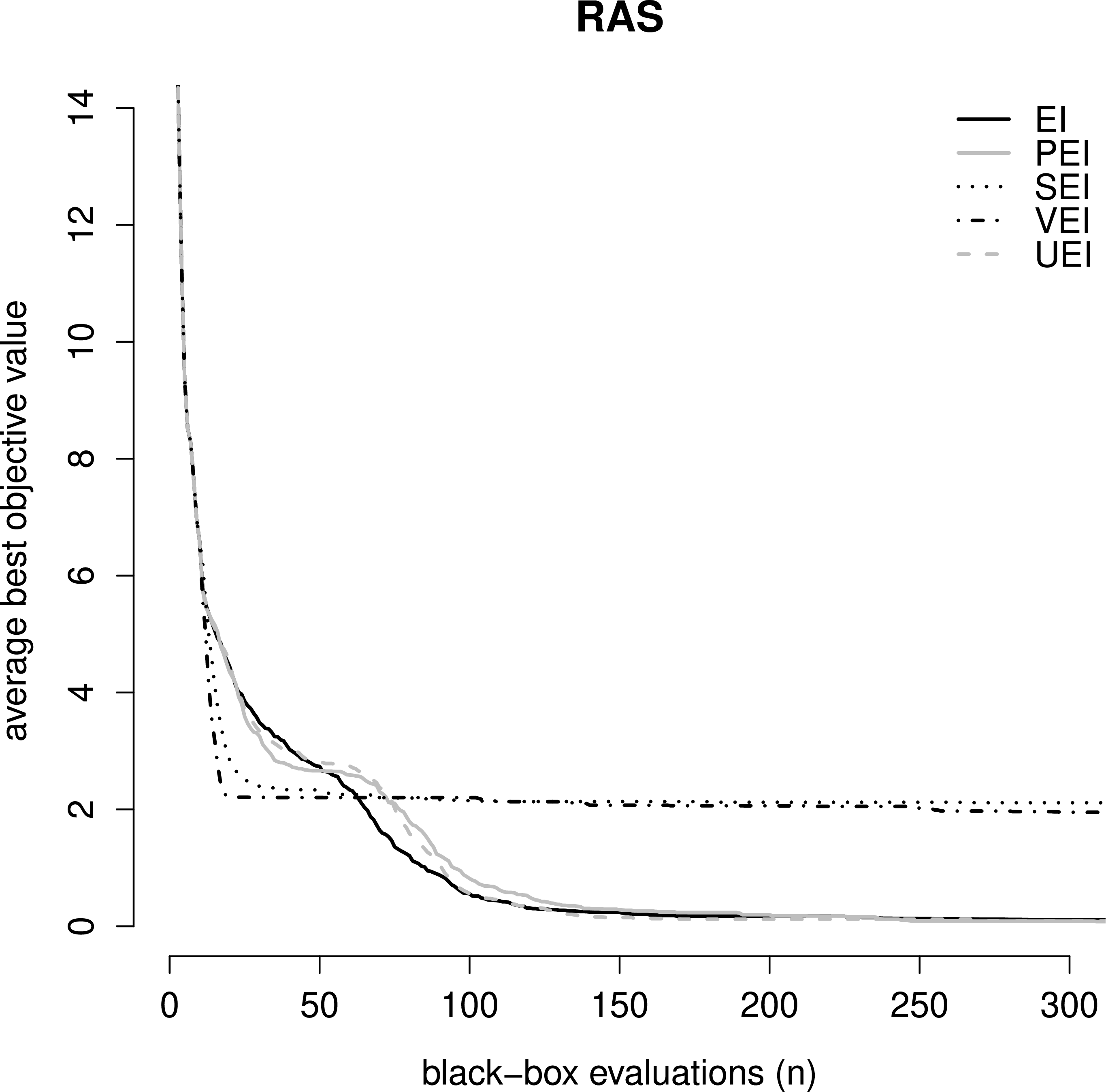}
    \includegraphics[scale = .13]{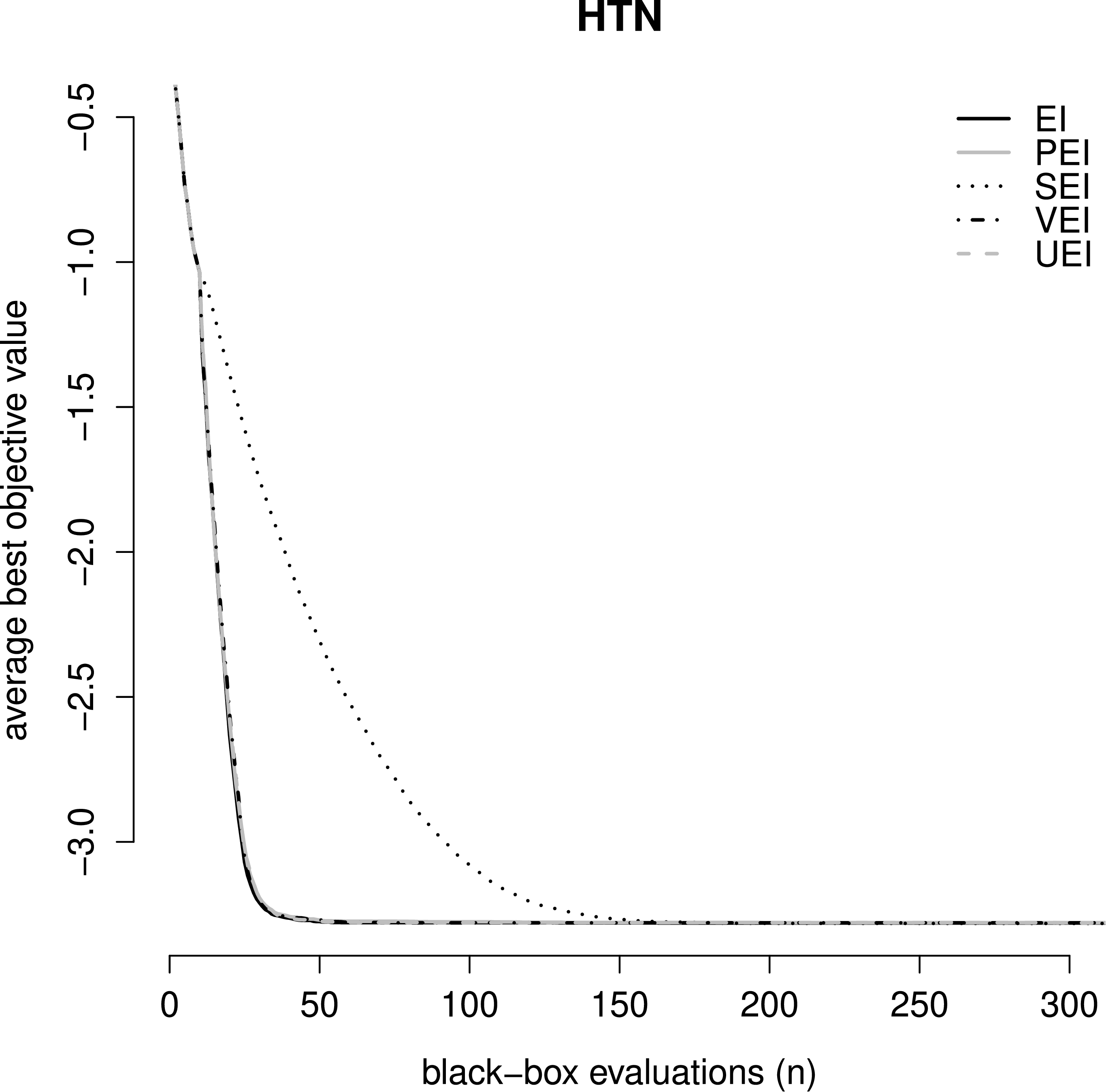}
    \end{tabular}
    \caption{The results of running 100 Monte Carlo repetitions, with random starting inputs, for the different test and acquisition functions. The plots show the average best objective function values found over 300 black-box iterations.}
    \label{fig:results2}
\end{figure}

For a given acquisition function, we solve each test problem by starting with an initial random sample of 10 inputs from a Latin hypercube design \citep{mcka:cono:beck:1979} over the input space, and then sequentially chose 490 more inputs based on our BO strategy. For each of the acquisition functions, we conduct 100 repetitions of a Monte Carlo experiment in order to quantify robustness and distribution of the solutions, as well as to understand under which scenarios a given acquisition function may (or may not) have difficulties or shortcomings in finding the global solution to the optimization problem. Table \ref{tab:solutions} and Figure \ref{fig:results2} capture the results of these different Monte Carlo experiments.

\begin{table}[htb]
\begin{tabular}{l c c c c c}\hline
Test & Acquisiton & Average Final & SD of Final & Best Final & Worst Final\\
Function & Function & Solution & Solution & Solution & Solution\\
\hline
\multirow{5}{*}{GRL} & EI & \bf{-0.867} & \bf{0.021} &  \bf{-0.869} & \bf{-0.663}\\
 & PEI  & -0.862 & 0.051 & \bf{-0.869} & -0.489\\
 & SEI  & -0.666 & 0.189 & \bf{-0.869} & -0.232\\
 & VEI  & -0.858 & 0.056 & \bf{-0.869} & -0.527\\
 & UEI  & -0.860 & 0.055 & \bf{-0.869} & -0.402\\
 \hline
 \multirow{5}{*}{ROS} & EI  & 0.002 & 0.005 & 2e-08 & 0.028\\
 & PEI  & 0.002 & 0.006 & 1e-08 & 0.039\\
 & SEI  & 0.103 & 0.136 & 8e-08 & 0.654\\
 & VEI  & 0.125 & 0.326 & 1e-06 & 2.83\\
 & UEI  & \bf{0.001} & \bf{0.002} & \bf{1e-10} & \bf{0.009}\\
 \hline
 \multirow{5}{*}{MOT} & EI  & -2.871 & 0.279 & -2.969 & -1.660\\
 & PEI  & -2.927 & 0.162 & -2.969 & -1.659\\
 & SEI  & -2.080 & 0.574 & -2.969 & -1.639\\
 & VEI  & -2.822 & 0.339 & -2.969 & -1.640\\
 & UEI  & \bf{-2.969} & \bf{3e-06} & \bf{-2.969} & -\bf{2.969}\\
\hline
 \multirow{5}{*}{ACY} & EI & 0.009 & 0.010 & 6e-05 & 0.044\\
 & PEI  & \bf{0.008} & \bf{0.010} & 8e-05& 0.041\\
 & SEI  & 0.009 & 0.010 & 6e-05 & 0.044\\
 & VEI  & 0.012 & 0.011 & 0.000 & 0.054\\
 & UEI  & 0.008 & 0.010 & \bf{4e-06} & \bf{0.039}\\
\hline
 \multirow{5}{*}{RAS} & EI & 0.094 & 0.488 & 1e-12 & \bf{4.000}\\
 & PEI  & 0.040 & 0.398 & 1e-10 & 3.981\\
 & SEI  & 2.103 & 1.481 & 2e-07 & 4.995\\
 & VEI  & 1.602 & 1.149 & 7e-07 & 4.995\\
 & UEI  & \bf{0.057} & \bf{0.431} & \bf{3e-12} & 4.000\\
 \hline
 \multirow{5}{*}{HTN} & EI & \bf{-3.280} & 0.059 & -3.322 & -3.137\\
 & PEI  & -3.279 & 0.058 & -3.322 & -3.193\\
 & SEI  & -3.281 & \bf{0.057} & -3.322 & -3.196\\
 & VEI  & -3.279 & 0.058 & \bf{-3.322} & \bf{-3.201}\\
 & UEI  & -3.279 & 0.058 & -3.322 & -3.201\\
\hline
\end{tabular}
\caption{The average, standard deviation (SD), best, and worst solutions found at the end of the 100 Monte Carlo experiments by each acquisition function on each optimization test function. Bolded values signify the best outcome in a given category for each test function.}
\label{tab:solutions}
\end{table}

\begin{figure}[htb]
    \centering
    \begin{tabular}{cc}
    \includegraphics[scale = .1]{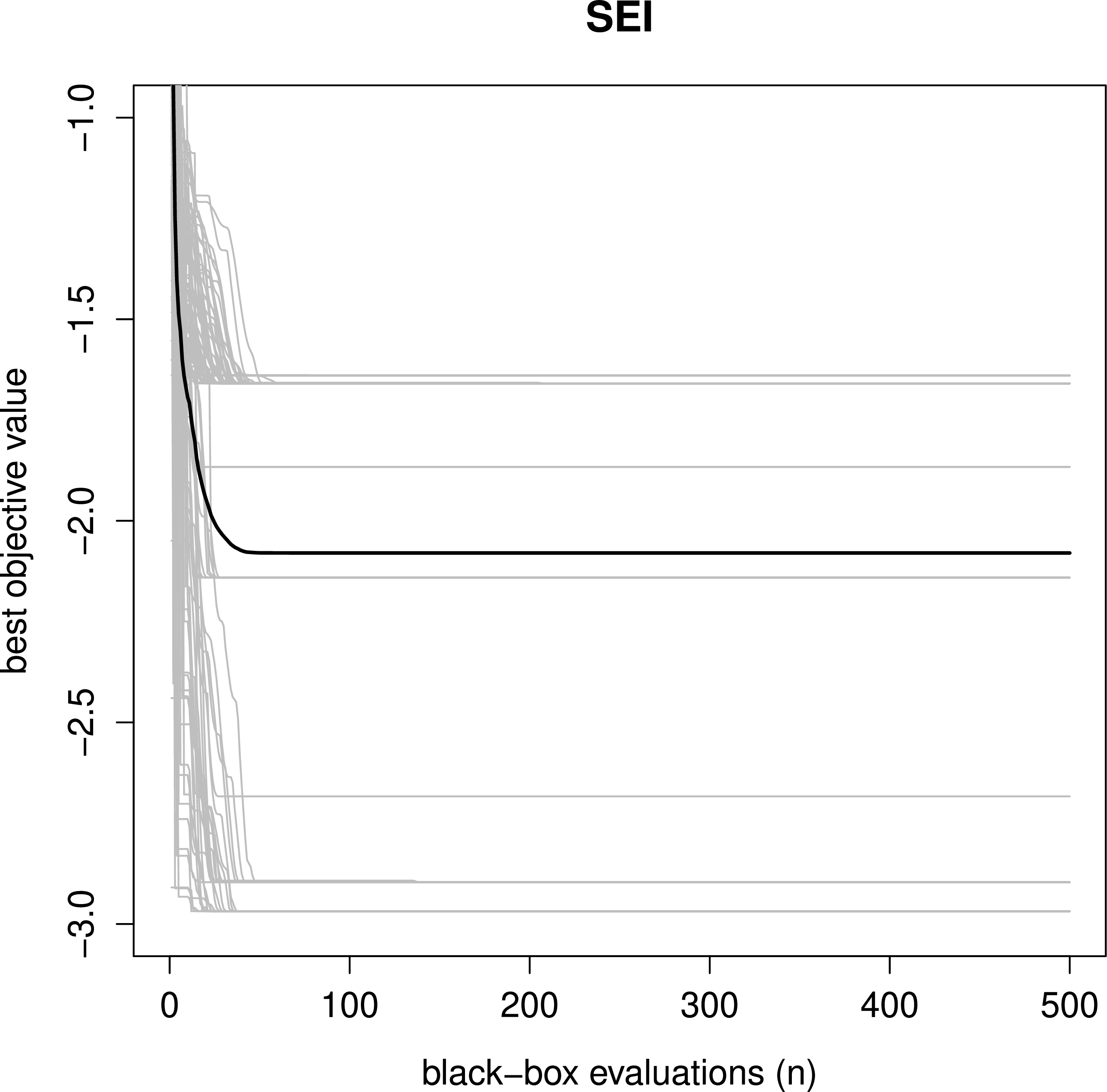}
    \includegraphics[scale = .1]{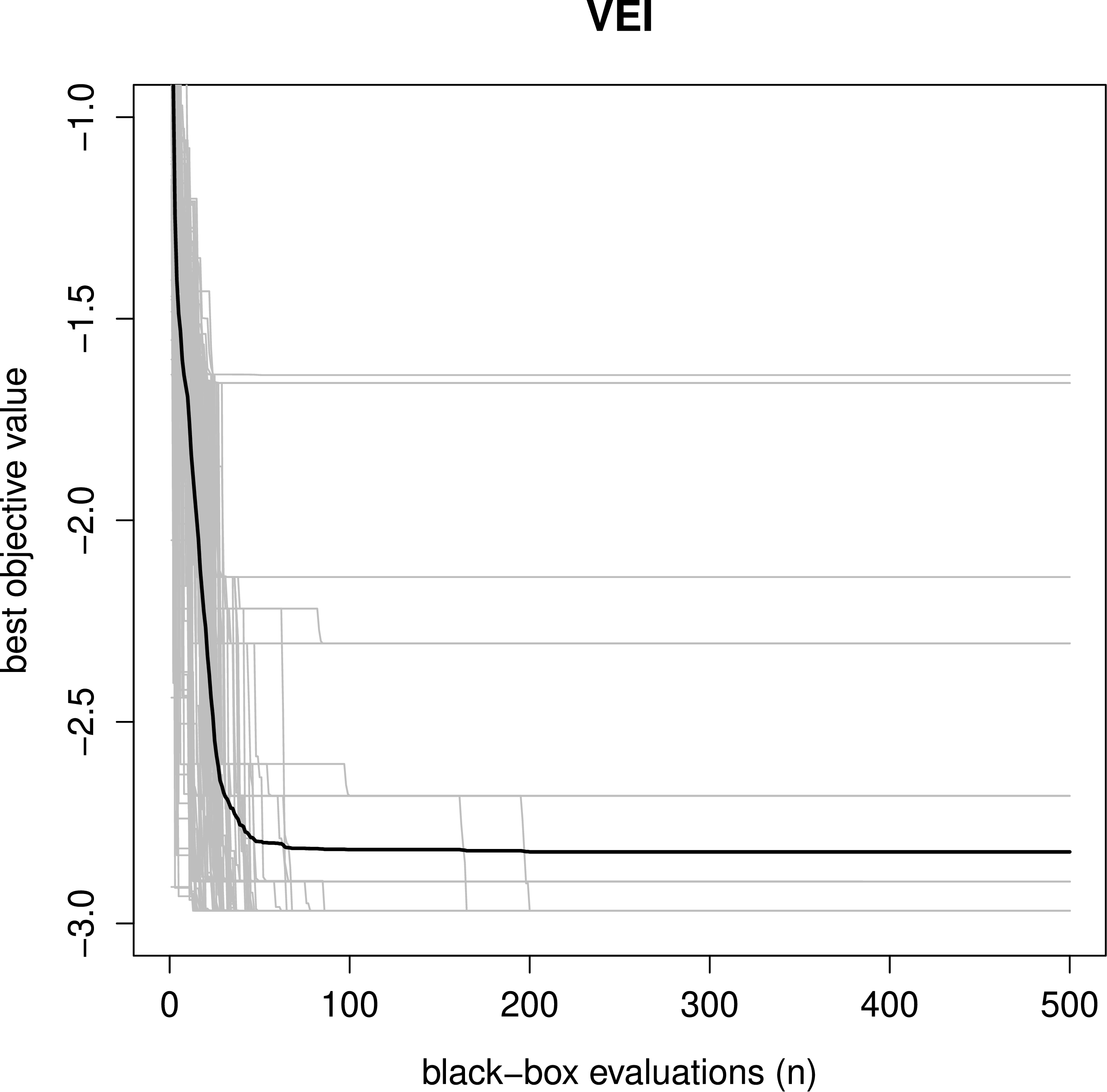}
    \includegraphics[scale = .1]{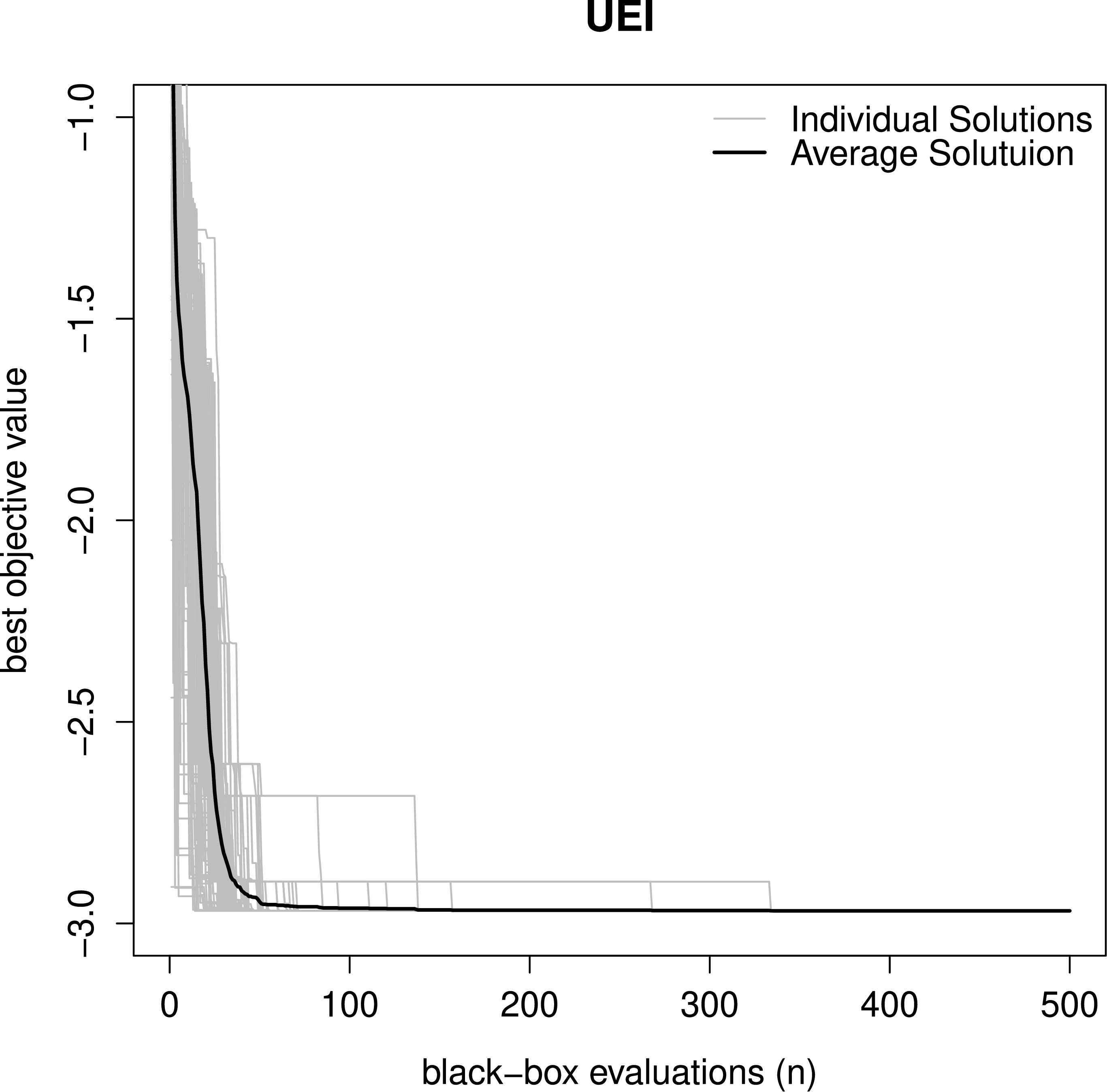}
    \end{tabular}
    \caption{A view of the performance of the SEI, VEI, and UEI acquisition functions for the 100 the Monte Carlo experiments associated with the MOT test function. Here, each grey line represents the best value found over the search by the BO algorithm during a single run of the Monte Carlo experiment while the black line represents the average of the grey lines.}
    \label{fig:individualruns}
\end{figure}

In general, for all six of the test functions, it appears that each acquisition function converges to the global solution of the optimization problem at least once (see the best final solution column of Table \ref{tab:solutions}), although some acquisition functions tended to find the global solution more reliably than others. Interestingly, the UEI acquisition function tended to do the best in all of the different performance categories of Table \ref{tab:solutions}), i.e., it had the best average final solution (3 out of 6 times), the smallest standard deviation of the final solution (3 out of 6 times), the best final solution (5 out of 6 times), and the smallest worst final solution (3 out of 6 times), over all of the different test functions. Furthermore, even when not the best acquisition function for a given test function, the UEI acquisition function tended to be as competitive as all of the other acquisition functions. These empirical results are indicative of the benefit of incorporating the variance of the improvement function into the acquisition function, but they also highlight the importance of incorporating that extra information efficiently into the acquisition function. The pitfalls of incorporating the variance of the improvement function into the acquisition as the SEI and VEI acquisition functions do was discussed in Section \ref{sec:varimprov}, and these issues readily manifest themselves in 4 of the 6 optimization test functions (the GRL, ROS, MOT, and RAS test functions). Investigating the shapes of the objective function surfaces (see \ref{append}), there does not appear to be a single distinguishing quality that leads to this poor performance, but rather, that the penalization for the variance in the improvement function leads the SEI and VEI acquisition functions to insufficiently  explore the input space. For example, if we investigate the individual Monte Carlo solutions associated with the MOT optimization function for the SEI, VEI, and UEI acquisition functions (Figure \ref{fig:individualruns}), we see that the SEI and VEI acquisition functions tended to have several runs that get stuck exploring around a local minima of the surface, while this does not tend to occur to the UEI acquisition function. The ROS function is unimodal but with a very shallow slope in a narrow valley around the minimum, and so being too focused on a local search can lead to slow movement within the valley and worse results for a fixed number of iterations.

On the other hand, there are clearly instances of when the the SEI and VEI acquisition functions do bring added value to the BO algorithm. For example, from Figure~\ref{fig:results2}, we see that for the ACY and RAS optimization test functions, the VEI and SEI acquisition functions were much quicker at minimizing the objective function with fewer runs than the other acquisition functions. Furthermore, the totality of the results in Table~\ref{tab:solutions} and Figure~\ref{fig:results2} seem to indicate that there that are potential strengths to try to borrow across all of the acquisition functions, further suggesting that a general family of acquisition functions is likely appropriate.

\subsection{Acquisition Function Family Performance}\label{sec:affp}
In this section, we tested the performance of BO using acquisition functions with different parameter values. The functions involved in the test include ROS, MOT, and RAS, with 1, 6, and 25 local minima, respectively. We use these functions to explore the behavior of various  parameter values under different difficulties of problems. For the selection of parameter sets, we try to keep some parameters fixed while varying others, because the parameters $w$, $u$, and $v$ all effect the influence of $\text{VI}(x)$ on the acquisition function. There are three groups of parameters involved in the test. The first group is to test only $w$. Here,  $u$ and $\beta$ are fixed to 0 (noting that the value of $v$ does not affect the setting in this case), and $w$ is set to 0, 1, 2, and 3 respectively. The second group of parameters tests how the value of $\beta$ and $v$ influence the performance of UEI. Here, $w$ and $u$ are set to 0 and 1, respectively, and the values of $\beta$ and $v$ take values from the Cartesian product  $\{-0.5, 0, 2\}\times\{0, 0.5, 1\}$. The last group of tests simultaneously adjusts $u$ and $v$, with $w$ and $\beta$ fixed to 1 and 2. The values of $u$ and $v$ take values from the Cartesian product $\{0, 0.5, 1\}\times\{0, 0.5, 1\}$. Notice that each set of test parameters contains EI as a reference baseline. 
	
\begin{figure}[htbp!]
    \centering
    \includegraphics[scale = .19]{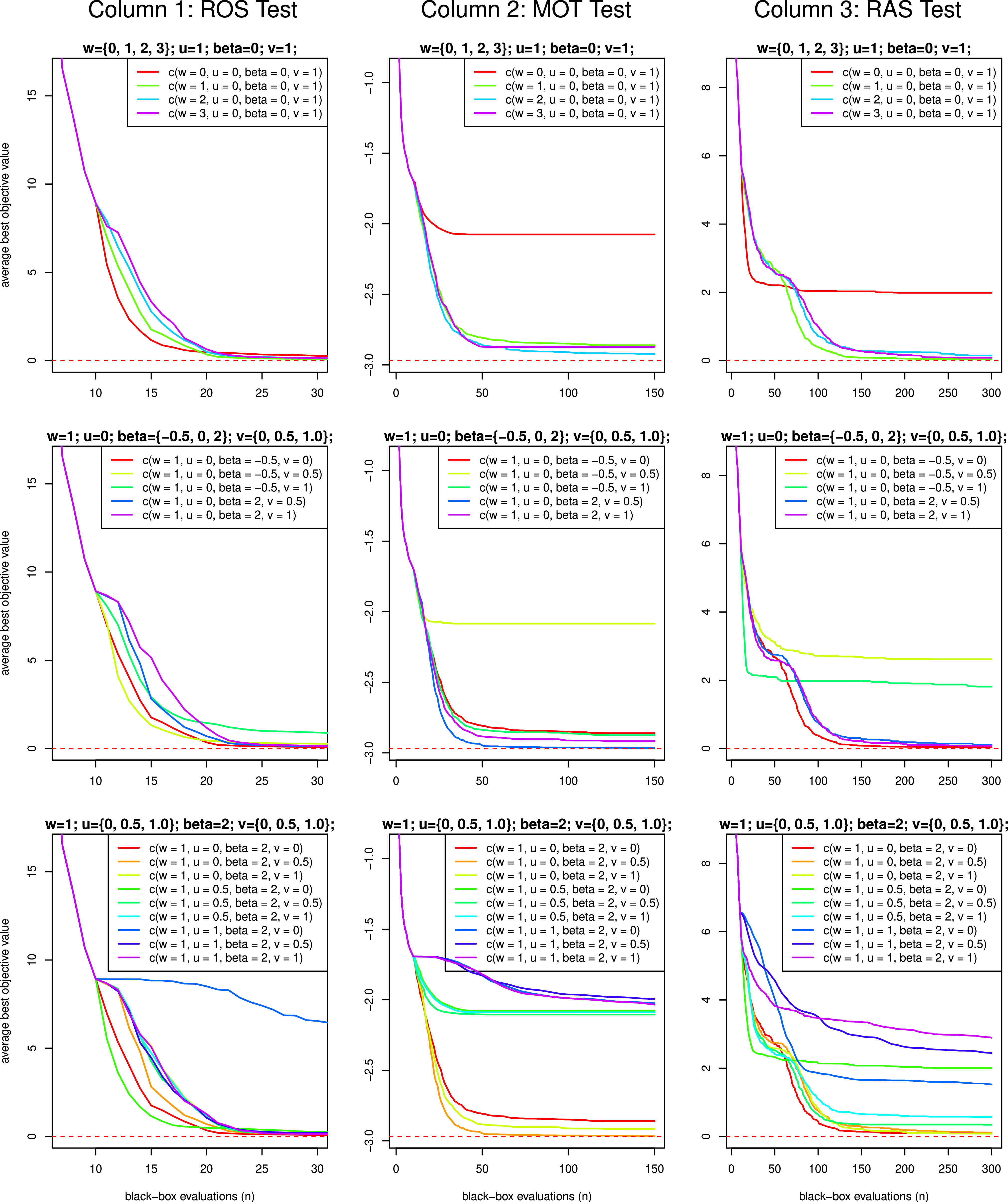}
    \caption{Performance of the BO algorithm under different parameter settings of the acquisition function family.}
    \label{fig:familyfunction}
\end{figure}

For the first set of parameters, only the value of $w$ is varied. The results of this part are shown in the first row of Figure~\ref{fig:familyfunction}. When $w=1$, it becomes the classic EI, which can be used as a reference; when $w=0$, the acquisition function becomes the probability of improvement (PI) \citep{kushner:1964}, and the algorithm will look for the position with a positive number in the surrogate function in each iteration, regardless of the size of the improvement. It can be seen that PI drops faster at the beginning in the ROS and RAS test functions, but does not converge in all the tests. When $w=2$, it becomes the classic PEI; since $\mathbb{E}(I^2(x))=\text{EI}(x)^2+\text{VI}(x)$, PEI favors locations with higher $\text{VI}(x)$. In the ROS and RAS function tests, it can be seen that PEI drops more slowly than EI at the beginning, but is more likely to find the global minima at the end. In the MOT function test, PEI dominates EI after multiple iterations. In other words, PEI more easily finds the global minimum than EI, and does not get stuck in local minima as much. When $w=3$, it can be seen in the ROS function that it drops more slowly than PEI, but finally finds the global minimum. Interestingly, in the MOT function test, the value of the lowest point does not go down like PEI. Instead, it gets stuck somewhere after about 50 iterations. That is to say, the algorithm will not guarantee that it will converge just because the value of $u$ is increased, and too much weight on VI will weaken the ability of the algorithm. See \cite{scho:welc:jone:1998} for more discussion of the form of $\E(I^w)$.

The second set of parameters explores how $\beta$ and $v$ influence the performance of VEI and UEI. The results are shown in the second row of Figure~\ref{fig:familyfunction}. We let $\beta$ and $v$ take values from the Cartesian product of $\{-0.5, 0, 2\}\times\{0, 0.5, 1\}$, and the acquisition function is equivalent to the classic EI when $\beta=0$ or $v=1$. Thus the tested parameter sets are effectively $\{0, 0\}$, $\{-0.5, 0.5\}$, $\{-0.5, 1\}$, $\{2, 0.5\}$ and $\{2, 1\}$. The algorithm is VEI when $\beta=-0.5$. VEI will penalize the location with higher variance. In the simple function ROS test, the algorithm will approach the lowest point faster than EI when $v=0.5$, but facing the more complex MOT and RAS functions, VEI quickly becomes stuck at a local minima; it is worth noting that when $v=1$, the algorithm is less likely to get stuck at the local minima, because after some iterations, there are more and more points used to generate the surrogate process, resulting in smaller and smaller $\text{VI}(x)$. When $0<\mbox{\text{VI}(x)}<1$, we have that $\mbox{\text{VI}(x)}<\mbox{\text{VI}(x)}^{1/2}$ resulting in a reduction in the penalty of $\text{VI}(x)$ for the acquisition function, which decreases the tendency of the algorithm to become stuck in local minima. The acquisition function is UEI when $\beta=2$. It can be seen from the unimodal ROS function that for either $v=0.5$ or $v=1$, UEI converges slower than the alternatives, as it puts more emphasis on global search than these comparators. However, when facing the more complex MOT and RAS functions, UEI is more capable of finding global minima. Interestingly when $v=0.5$, the performance in the MOT function test is significantly better than any other parameter choices and is the only one that converges every time. Compared with $v=1$, the performance of $v=0.5$ is better mainly because the reward for $\text{VI}(x)$ is greater in the long run as $\text{VI}(x)$ decreases below $1$.
	
The last comparison fixes $w=1$ and $\beta=2$ and explores the influence of adjusting $u$ and $v$. When $u=0$ and $v=0.5$ we again obtain UEI. Meanwhile, when $u=0.5$ and $v=0$, the acquisition function is reduced to SEI. When both $u$ and $v$ are 0, the acquisition function returns to the classic EI. In these examples, SEI ($v=0$ and $u=0.5$) displays similar behavior as VEI, converging the fastest for the simple ROS function, but quickly becoming stuck in local minima when for the more complex MOT and RAS functions. It is worth noting that $u=1$ makes SEI converge even slower, because the $\mbox{\text{VI}(x)}>1$ in the beginning, making the acquisition function quite risk averse. At each iteration, the algorithm will prefer points that are very close to the known points because of the $\text{VI}(x)$ penalty, making searches extremely localized. When $u\neq0$, all parameter sets perform poorly on the complex function tests. $u\neq0$ penalizes risk, while $v\neq0$ rewards risk, and when they are both non-zero, the dynamic can be complicated and less effective. 

Overall we see the best performance when $u=0$, especially when paired with $v=0.5$, which results in the UEI acquisition function when $\beta=2$. Most of the acquisition functions in this family succeed in finding the global minimum for simple functions such as ROS. Acquisition functions that penalize $\text{VI}(x)$, such as VEI and SEI, usually require fewer iterations to converge on simpler functions. For objective functions that contain multiple local minima, the acquisition functions that reward $\text{VI}(x)$ are more likely to find the global minimum.

\section{Discussion}
\label{sec:disc}
This paper considers a range of acquisition functions as part of a larger family, in order to explore the behavior of functions in the family as parameter values shift. This family includes the existing acquisition functions EI, PEI, SEI, and VEI, and it also includes a newly proposed UEI. A simulation study illustrates the effect of tuning different parameters for this family.

The simulation results show that penalization of the variance results in faster convergence to local minima, while rewarding variance improves global search, resulting in improved convergence to the global minimum on complex multimodal objective functions. Thus, the best choice of acquisition function will depend on the particular objective function. If some information about the objective function is available, analyses such as the one in this paper can be used to guide a more effective choice of acquisition function.


\backmatter








\section*{Statements and Declarations}
The authors declare that there are no conflicts of interests nor competing interests. Additionally, there is no data associated with this publication. 

\begin{appendices}

\section{Test Functions for Optimization}\label{append}
Further descriptions and implementations of the test functions in this section can be found in \cite{simulationlib}. Here we list the functional forms of the six test functions we examined in this paper, along with their respective input domains, and plots when applicable. Recall that the objective here is to minimize $f(x)$ subject to its input domain $x$.\\

\textbf{Gramacy and Lee Function}
\begin{align}
    f(x) &= \frac{\sin(10\pi x)}{2x} + (x-1)^4\\\nonumber
    &x\in[0.5,2.5]
\end{align}

\begin{figure}[htb]
    \centering
    \includegraphics[scale = .2]{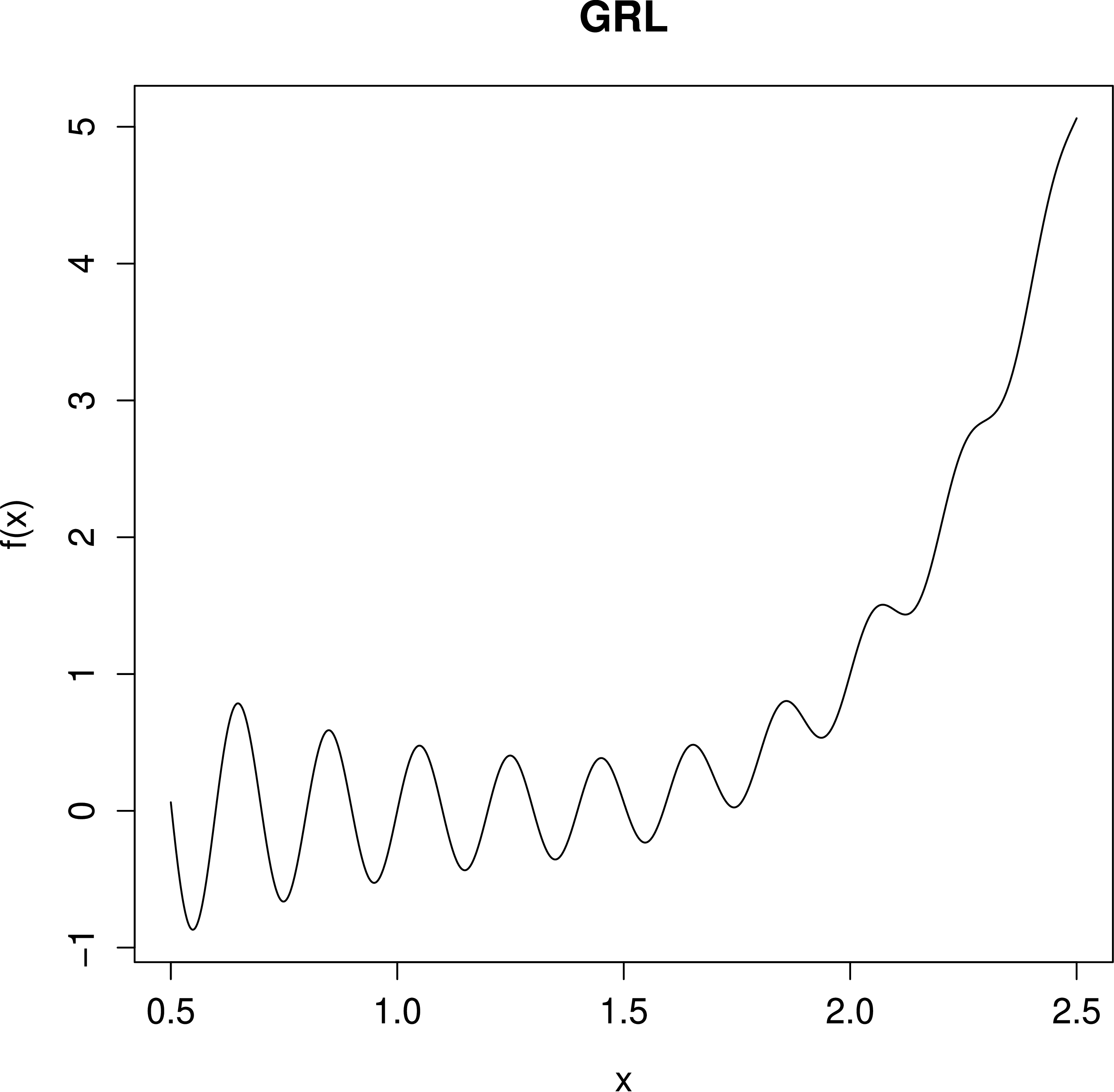}
    \caption{The Gramacy and Lee optimization test function.}
    \label{fig:HRB}
\end{figure}

\textbf{Rosenbrock Function}
\begin{align}
    f(x) &= 100(x_2-x_1^2)^2 + (x_1-1)^2\\\nonumber
    &x_1,x_2\in[-2,2]
\end{align}

\begin{figure}[htb]
    \centering
    \begin{tabular}{cc}
    \includegraphics[scale = .14]{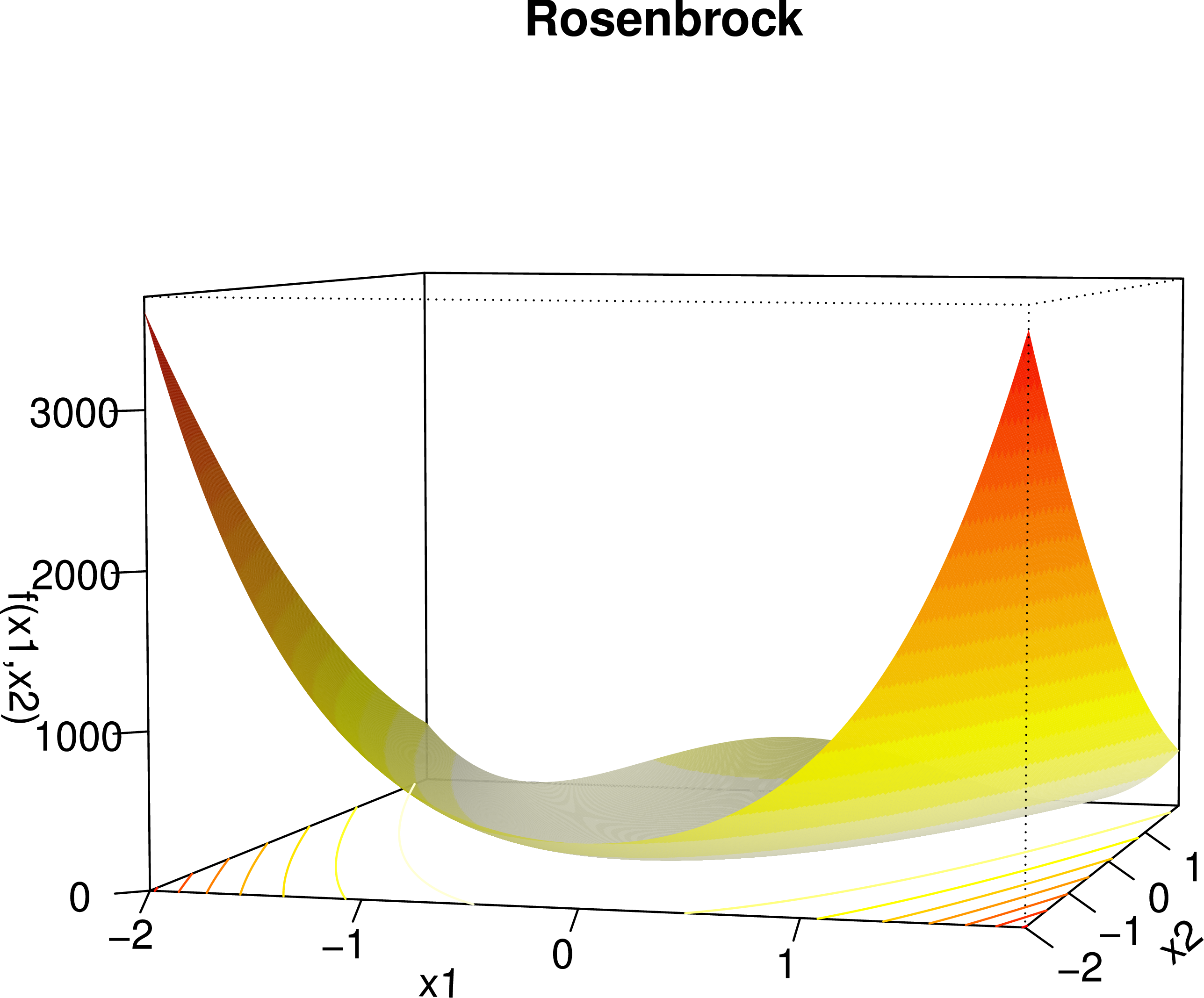}
    \includegraphics[scale = .14]{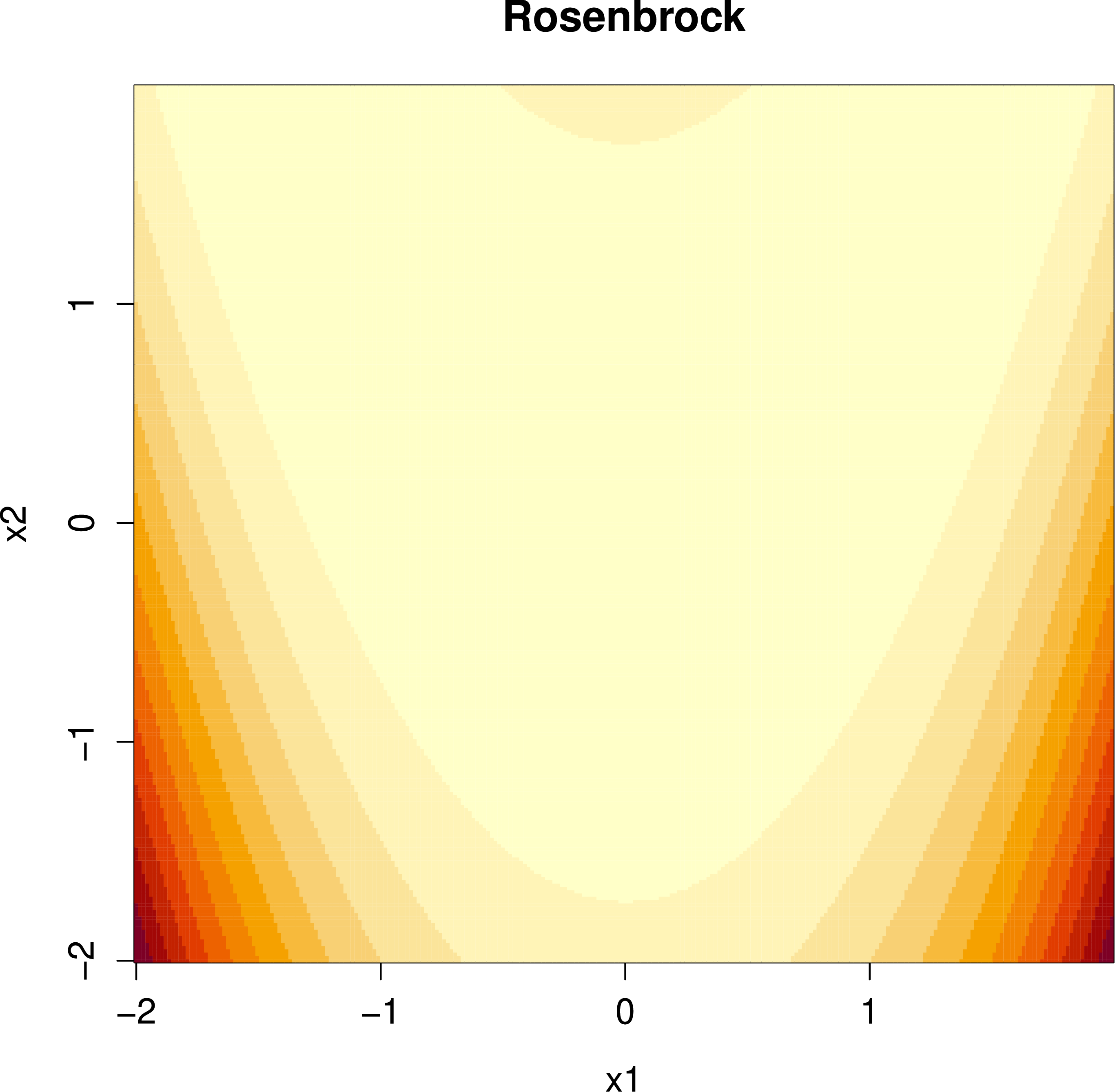}
    \end{tabular}
    \caption{The objective function surface for the Rosenbrock test function (left) and its associated contour plot (right).}
    \label{fig:ROS}
\end{figure}

\textbf{Modified Townsend Function}
\begin{align}
    f(x) &=-[\cos((x_1 - .1)x_2)]^2- x_1 \sin (3 x_1 + x_2)\\\nonumber
        &x_1,x_2\in[-2,2]
\end{align}

\begin{figure}[htb]
    \centering
    \begin{tabular}{cc}
    \includegraphics[scale = .14]{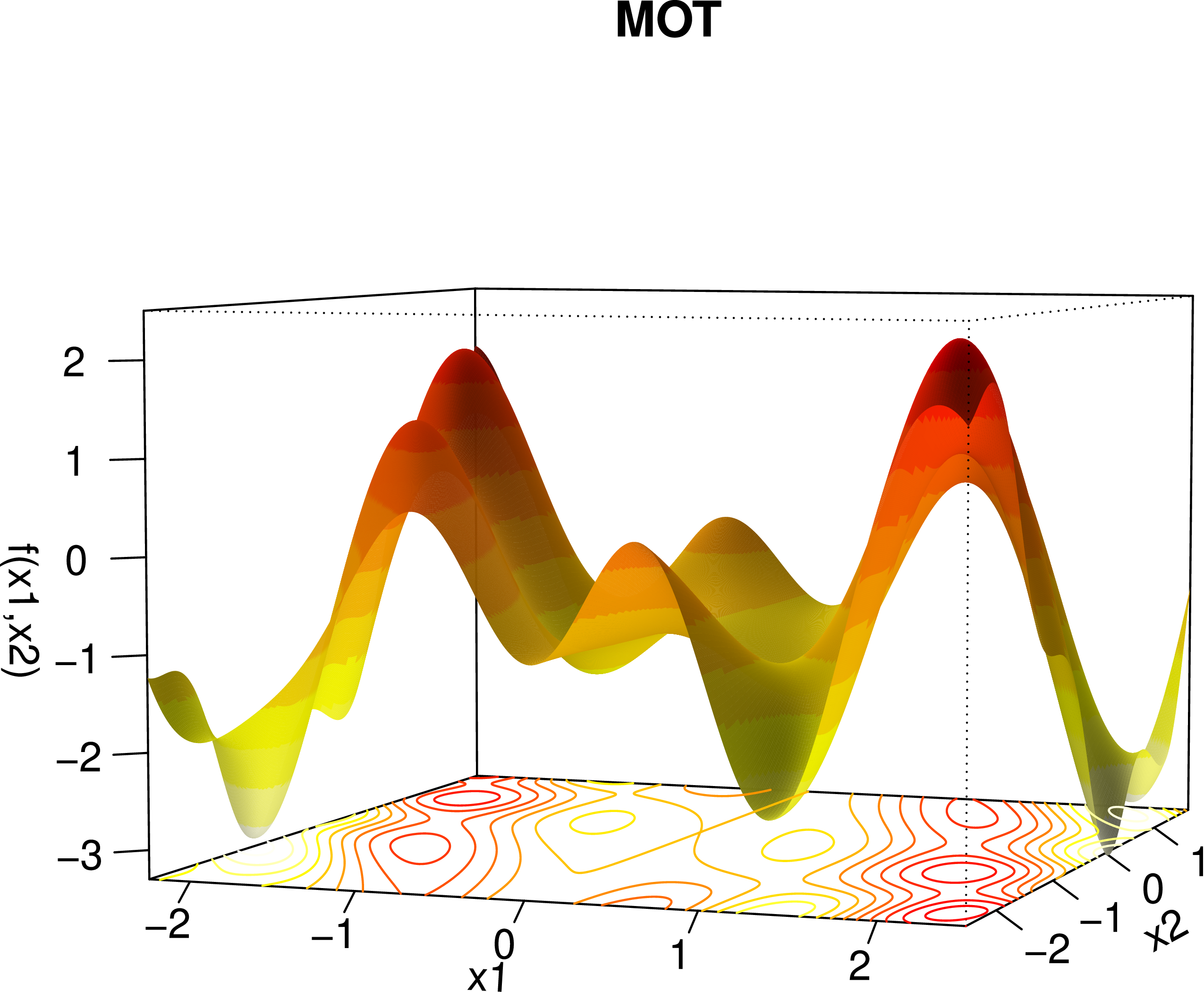}
    \includegraphics[scale = .14]{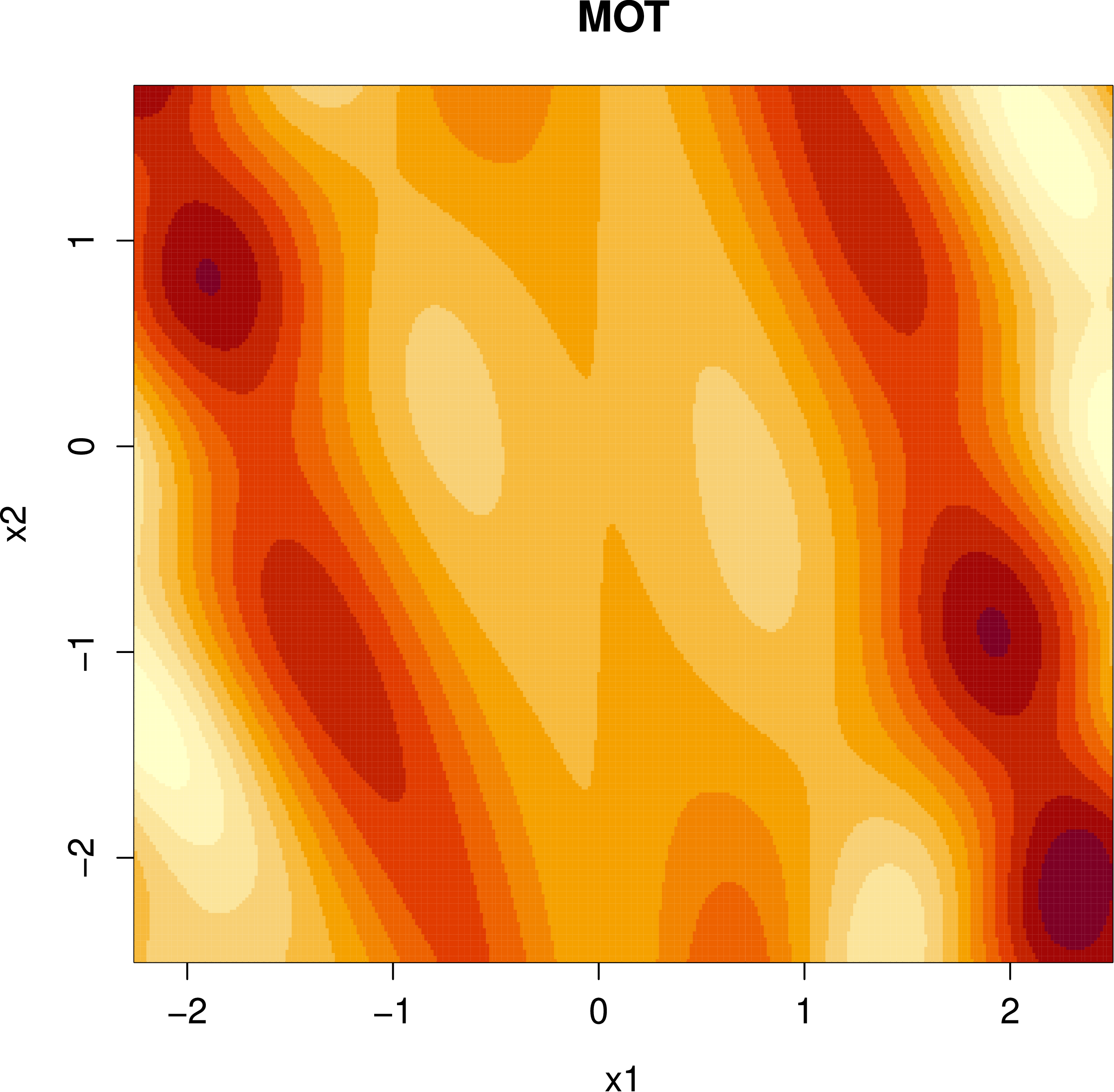}
    \end{tabular}
    \caption{The objective function surface for the Modified Townsend test function (left) and its associated contour plot (right).}
    \label{fig:MOT}
\end{figure}

\textbf{Ackley Function}
\begin{align}
    f(x) &=-20\exp\left(-0.2\sqrt{\frac{1}{2}\sum_{i=1}^2x_i^2}\right)-\exp\left(\frac{1}{2}\sum_{i=1}^2\cos(2\pi x_i)\right) + 20 + \exp(1)\\\nonumber
        &x_1,x_2\in[-2,2]
\end{align}

\begin{figure}[htb]
    \centering
    \begin{tabular}{cc}
    \includegraphics[scale = .14]{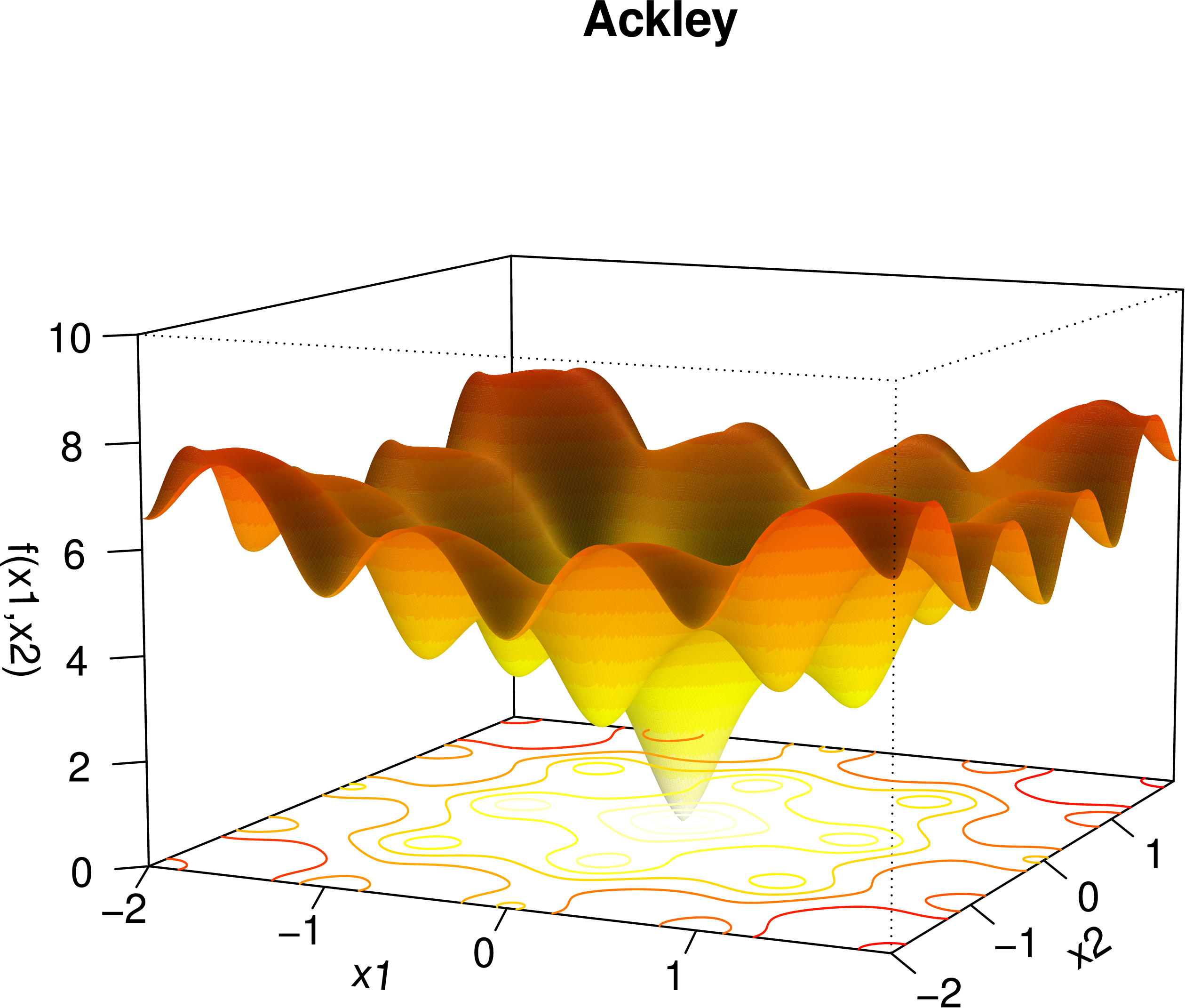}
    \includegraphics[scale = .14]{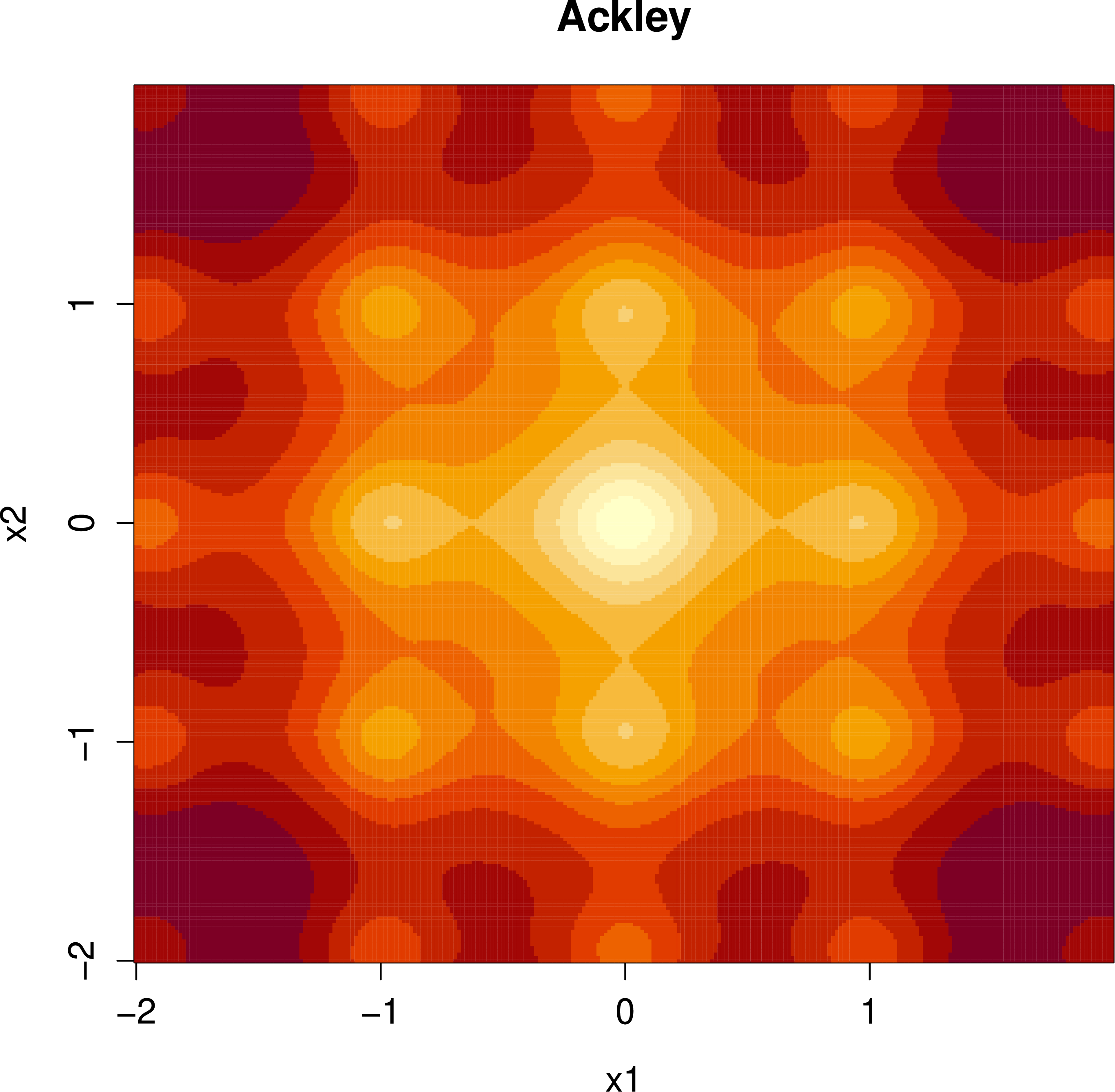}
    \end{tabular}
    \caption{The objective function surface for the Ackley test function (left) and its associated contour plot (right).}
    \label{fig:ACK}
\end{figure}

\textbf{Rastrigin Function}
\begin{align}
    f(x) &= 20 + \sum_{i=1}^2\left(x_i^2 - 10\cos(2\pi x_i)\right)\\\nonumber
        &x_1,x_2\in[-2,2]
\end{align}

\begin{figure}[htb]
    \centering
    \begin{tabular}{cc}
    \includegraphics[scale = .14]{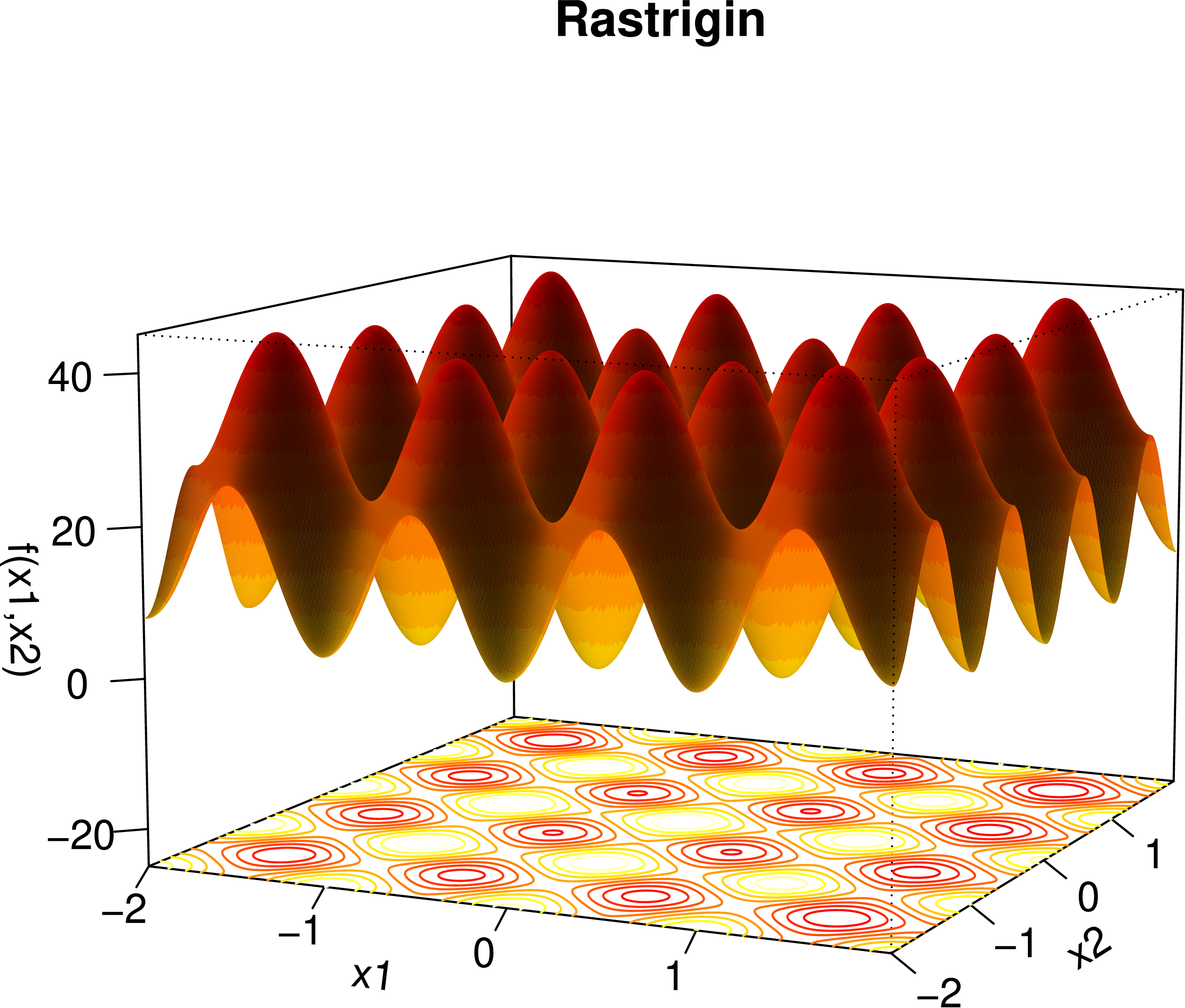}
    \includegraphics[scale = .14]{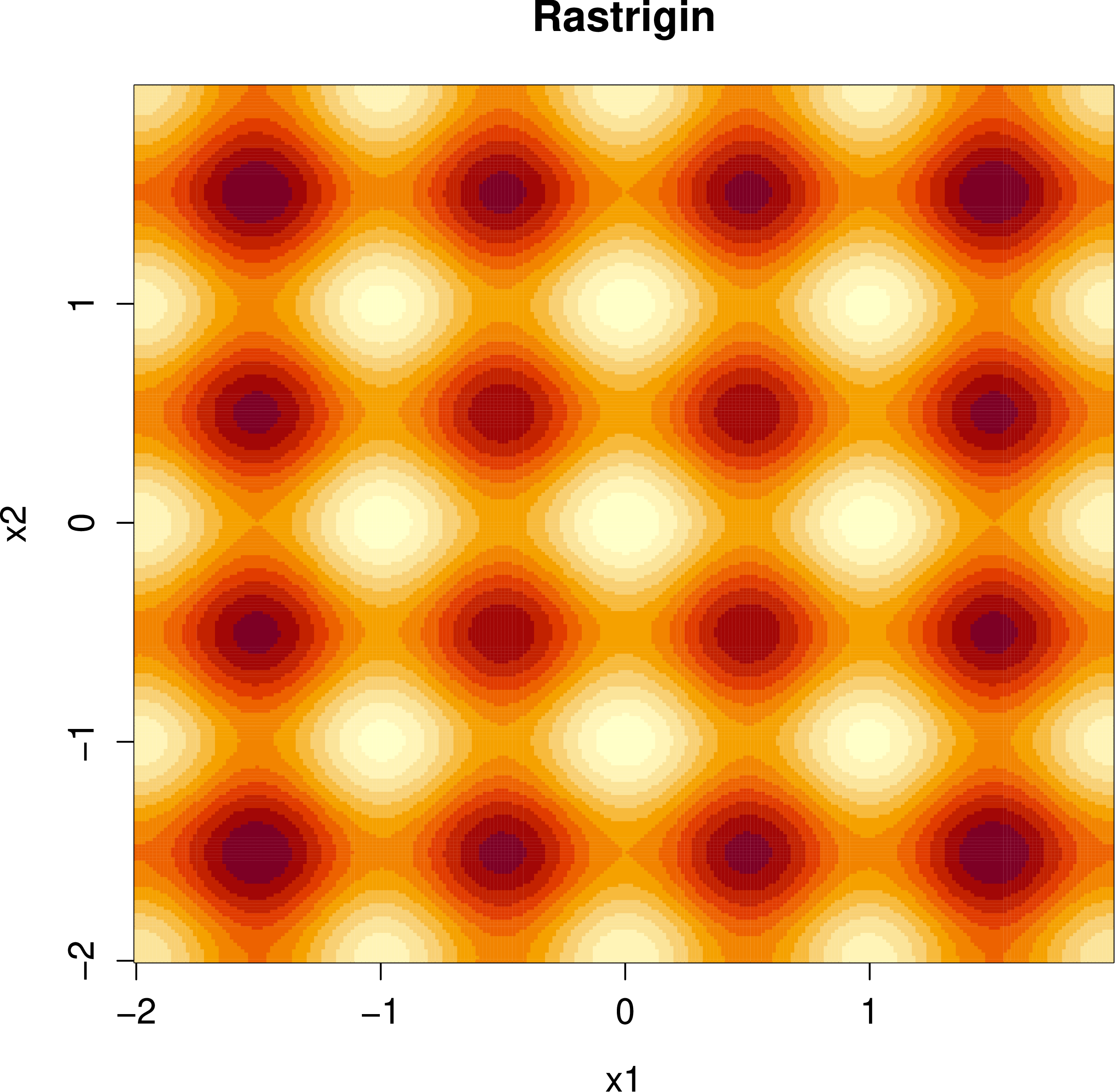}
    \end{tabular}
    \caption{The objective function surface for the Rastrigin test function (left) and its associated contour plot (right).}
    \label{fig:RAS}
\end{figure}

\textbf{Hartman Function}
\begin{align}
    f(x) =-\sum_{i=1}^4\alpha_i\exp\left(-\sum_{j=1}^6A_{ij}(x_j-P_{ij})^2\right),
\end{align}
where $\alpha=(1,1.2,3,3.2)^T$, 
\begin{align}
    A = \begin{pmatrix}
        10 & 3 & 17 & 3.5 & 1.7 & 8\\
        0.05 & 10 & 17 & 0.1 & 8 & 14\\
        3 & 3.5 & 1.7 & 10 & 17 & 8\\
        17 & 8 & 0.05 & 10 & 0.1 & 14
    \end{pmatrix},
\end{align}
and
\begin{align}
    P = 10^{-4}\begin{pmatrix}
        1312 & 1696 & 5569 & 124 & 8283 & 5886\\
        2329 & 4135 & 8307 & 3736 & 1004 & 9991\\
        2348 & 1451 & 3522 & 2883 & 3047 & 6650\\
        4047 & 8828 & 8732 & 5743 & 1091 & 381
    \end{pmatrix},
\end{align}
where $x_i\in(0,1)$ for all $i=1,...,6$.

\end{appendices}

\bibliography{bibs/bibliography}


\begin{thebibliography}{20}
\ifx \bisbn   \undefined \def \bisbn  #1{ISBN #1}\fi
\ifx \binits  \undefined \def \binits#1{#1}\fi
\ifx \bauthor  \undefined \def \bauthor#1{#1}\fi
\ifx \batitle  \undefined \def \batitle#1{#1}\fi
\ifx \bjtitle  \undefined \def \bjtitle#1{#1}\fi
\ifx \bvolume  \undefined \def \bvolume#1{\textbf{#1}}\fi
\ifx \byear  \undefined \def \byear#1{#1}\fi
\ifx \bissue  \undefined \def \bissue#1{#1}\fi
\ifx \bfpage  \undefined \def \bfpage#1{#1}\fi
\ifx \blpage  \undefined \def \blpage #1{#1}\fi
\ifx \burl  \undefined \def \burl#1{\textsf{#1}}\fi
\ifx \doiurl  \undefined \def \doiurl#1{\url{https://doi.org/#1}}\fi
\ifx \betal  \undefined \def \betal{\textit{et al.}}\fi
\ifx \binstitute  \undefined \def \binstitute#1{#1}\fi
\ifx \binstitutionaled  \undefined \def \binstitutionaled#1{#1}\fi
\ifx \bctitle  \undefined \def \bctitle#1{#1}\fi
\ifx \beditor  \undefined \def \beditor#1{#1}\fi
\ifx \bpublisher  \undefined \def \bpublisher#1{#1}\fi
\ifx \bbtitle  \undefined \def \bbtitle#1{#1}\fi
\ifx \bedition  \undefined \def \bedition#1{#1}\fi
\ifx \bseriesno  \undefined \def \bseriesno#1{#1}\fi
\ifx \blocation  \undefined \def \blocation#1{#1}\fi
\ifx \bsertitle  \undefined \def \bsertitle#1{#1}\fi
\ifx \bsnm \undefined \def \bsnm#1{#1}\fi
\ifx \bsuffix \undefined \def \bsuffix#1{#1}\fi
\ifx \bparticle \undefined \def \bparticle#1{#1}\fi
\ifx \barticle \undefined \def \barticle#1{#1}\fi
\bibcommenthead
\ifx \bconfdate \undefined \def \bconfdate #1{#1}\fi
\ifx \botherref \undefined \def \botherref #1{#1}\fi
\ifx \url \undefined \def \url#1{\textsf{#1}}\fi
\ifx \bchapter \undefined \def \bchapter#1{#1}\fi
\ifx \bbook \undefined \def \bbook#1{#1}\fi
\ifx \bcomment \undefined \def \bcomment#1{#1}\fi
\ifx \oauthor \undefined \def \oauthor#1{#1}\fi
\ifx \citeauthoryear \undefined \def \citeauthoryear#1{#1}\fi
\ifx \endbibitem  \undefined \def \endbibitem {}\fi
\ifx \bconflocation  \undefined \def \bconflocation#1{#1}\fi
\ifx \arxivurl  \undefined \def \arxivurl#1{\textsf{#1}}\fi
\csname PreBibitemsHook\endcsname

\bibitem[\protect\citeauthoryear{Gramacy}{2020}]{gramacy:2020}
\begin{bbook}
\bauthor{\bsnm{Gramacy}, \binits{R.B.}}:
\bbtitle{Surrogates: Gaussian Process Modeling, Design, and Optimization for
  the Applied Sciences}.
\bpublisher{Chapman \& Hall/CRC}, \blocation{???}
(\byear{2020})
\end{bbook}
\endbibitem

\bibitem[\protect\citeauthoryear{Pourmohamad and
  Lee}{2021}]{pourmohamad:book:2021}
\begin{bbook}
\bauthor{\bsnm{Pourmohamad}, \binits{T.}},
\bauthor{\bsnm{Lee}, \binits{H.K.H.}}:
\bbtitle{{B}ayesian Optimization with Application to Computer Experiments}.
\bpublisher{Springer}, \blocation{???}
(\byear{2021})
\end{bbook}
\endbibitem

\bibitem[\protect\citeauthoryear{Mockus et~al.}{1978}]{mockus:1978}
\begin{barticle}
\bauthor{\bsnm{Mockus}, \binits{J.}},
\bauthor{\bsnm{Tiesis}, \binits{V.}},
\bauthor{\bsnm{Zilinskas}, \binits{A.}}:
\batitle{The application of bayesian methods for seeking the extrenum}.
\bjtitle{Towards Global Optimization}
\bvolume{2},
\bfpage{117}--\blpage{129}
(\byear{1978})
\end{barticle}
\endbibitem

\bibitem[\protect\citeauthoryear{Brochu et~al.}{2010}]{brochu:2010}
\begin{botherref}
\oauthor{\bsnm{Brochu}, \binits{E.}},
\oauthor{\bsnm{Cora}, \binits{V.M.}},
\oauthor{\bsnm{Freitas}, \binits{N.}}:
A tutorial on bayesian optimization of expensive cost functions, with
  application to active user modeling and hierarchical reinforcement learning.
Technical Report 1012.2599,
arXiv
(2010)
\end{botherref}
\endbibitem

\bibitem[\protect\citeauthoryear{Conn
  et~al.}{2009}]{conn:scheinberg:vicente:2009}
\begin{bbook}
\bauthor{\bsnm{Conn}, \binits{A.R.}},
\bauthor{\bsnm{Scheinberg}, \binits{K.}},
\bauthor{\bsnm{Vicente}, \binits{L.N.}}:
\bbtitle{Introduction to Derivative-Free Optimization}.
\bpublisher{SIAM},
\blocation{Philadelphia}
(\byear{2009})
\end{bbook}
\endbibitem

\bibitem[\protect\citeauthoryear{Schonlau et~al.}{1998}]{scho:welc:jone:1998}
\begin{bchapter}
\bauthor{\bsnm{Schonlau}, \binits{M.}},
\bauthor{\bsnm{Jones}, \binits{D.R.}},
\bauthor{\bsnm{Welch}, \binits{W.J.}}:
\bctitle{Global versus local search in constrained optimization of computer
  models}.
In: \bbtitle{New Developments and Applications in Experimental Design}.
\bsertitle{IMS Lecture Notes - Monograph Series},
pp. \bfpage{11}--\blpage{25}
(\byear{1998})
\end{bchapter}
\endbibitem

\bibitem[\protect\citeauthoryear{Taddy et~al.}{2009}]{tadd:lee:gray:grif:2009}
\begin{barticle}
\bauthor{\bsnm{Taddy}, \binits{M.}},
\bauthor{\bsnm{Lee}, \binits{H.K.H.}},
\bauthor{\bsnm{Gray}, \binits{G.A.}},
\bauthor{\bsnm{Griffin}, \binits{J.D.}}:
\batitle{Bayesian guided pattern search for robust local optimization}.
\bjtitle{Technometrics}
\bvolume{51},
\bfpage{389}--\blpage{401}
(\byear{2009})
\end{barticle}
\endbibitem

\bibitem[\protect\citeauthoryear{Srinivas et~al.}{2010}]{srinivas:2010}
\begin{bchapter}
\bauthor{\bsnm{Srinivas}, \binits{N.}},
\bauthor{\bsnm{Krause}, \binits{A.}},
\bauthor{\bsnm{Kakade}, \binits{S.}},
\bauthor{\bsnm{Seeger}, \binits{M.}}:
\bctitle{Gaussian process optimization in the bandit setting: No regret and
  experimental design}.
In: \bbtitle{Proceedings of the 27th International Conference on Machine
  Learning}.
\bsertitle{ICML 2010}
(\byear{2010})
\end{bchapter}
\endbibitem

\bibitem[\protect\citeauthoryear{Snoek et~al.}{2012}]{snoek:2012}
\begin{bchapter}
\bauthor{\bsnm{Snoek}, \binits{J.}},
\bauthor{\bsnm{Larochelle}, \binits{H.}},
\bauthor{\bsnm{Adams}, \binits{R.P.}}:
\bctitle{Practical bayesian optimization of machine learning algorithms}.
In: \beditor{\bsnm{Pereira}, \binits{F.}},
\beditor{\bsnm{Burges}, \binits{C.J.C.}},
\beditor{\bsnm{Bottou}, \binits{L.}},
\beditor{\bsnm{Weinberger}, \binits{K.Q.}} (eds.)
\bbtitle{Advances in Neural Information Processing Systems 25},
pp. \bfpage{2951}--\blpage{2959}.
\bpublisher{Curran Associates, Inc.}, \blocation{???}
(\byear{2012})
\end{bchapter}
\endbibitem

\bibitem[\protect\citeauthoryear{Henning and
  Schuler}{2012}]{henning:shuler:2012}
\begin{barticle}
\bauthor{\bsnm{Henning}, \binits{P.}},
\bauthor{\bsnm{Schuler}, \binits{C.J.}}:
\batitle{Entropy search for information-efficient global optimization}.
\bjtitle{Journal of Machine Learning Research}
\bvolume{13},
\bfpage{1809}--\blpage{1837}
(\byear{2012})
\end{barticle}
\endbibitem

\bibitem[\protect\citeauthoryear{Hern\'{a}ndez-Lobato
  et~al.}{2014}]{hernandez:lobato:2014}
\begin{bchapter}
\bauthor{\bsnm{Hern\'{a}ndez-Lobato}, \binits{J.M.}},
\bauthor{\bsnm{Hoffman}, \binits{M.W.}},
\bauthor{\bsnm{Ghahramani}, \binits{Z.}}:
\bctitle{Predictive entropy search for efficient global optimization of
  black-box functions}.
In: \beditor{\bsnm{Ghahramani}, \binits{Z.}},
\beditor{\bsnm{Welling}, \binits{M.}},
\beditor{\bsnm{Cortes}, \binits{C.}},
\beditor{\bsnm{Lawrence}, \binits{N.}},
\beditor{\bsnm{Weinberger}, \binits{K.Q.}} (eds.)
\bbtitle{Advances in Neural Information Processing Systems}
vol. \bseriesno{27}.
\bpublisher{Curran Associates}, \blocation{???}
(\byear{2014})
\end{bchapter}
\endbibitem

\bibitem[\protect\citeauthoryear{Jones
  et~al.}{1998}]{jones:schonlau:welch:1998}
\begin{barticle}
\bauthor{\bsnm{Jones}, \binits{D.R.}},
\bauthor{\bsnm{Schonlau}, \binits{M.}},
\bauthor{\bsnm{Welch}, \binits{W.J.}}:
\batitle{Efficient global optimization of expensive black box functions}.
\bjtitle{Journal of Global Optimization}
\bvolume{13},
\bfpage{455}--\blpage{492}
(\byear{1998})
\end{barticle}
\endbibitem

\bibitem[\protect\citeauthoryear{No\`{e} and Husmeier}{2019}]{noe:2019}
\begin{botherref}
\oauthor{\bsnm{No\`{e}}, \binits{U.}},
\oauthor{\bsnm{Husmeier}, \binits{D.}}:
On a new improvement-based acquisition function for {B}ayesian optimization.
Technical Report 1808.06918,
arXiv
(2019)
\end{botherref}
\endbibitem

\bibitem[\protect\citeauthoryear{Marisu and Pun}{2023}]{marisu:2021}
\begin{barticle}
\bauthor{\bsnm{Marisu}, \binits{G.P.}},
\bauthor{\bsnm{Pun}, \binits{C.S.}}:
\batitle{Bayesian estimation and optimization for learning sequential
  regularized portfolios}.
\bjtitle{SIAM Journal on Financial Mathematics}
\bvolume{1},
\bfpage{1894}--\blpage{1899}
(\byear{2023})
\end{barticle}
\endbibitem

\bibitem[\protect\citeauthoryear{Frazier et~al.}{2008}]{frazier:2008}
\begin{barticle}
\bauthor{\bsnm{Frazier}, \binits{P.I.}},
\bauthor{\bsnm{Powell}, \binits{W.B.}},
\bauthor{\bsnm{Dayanik}, \binits{S.}}:
\batitle{A knowledge-gradient policy for sequential information collection}.
\bjtitle{SIAM Journal of Control Optimization}
\bvolume{47}(\bissue{5}),
\bfpage{2410}--\blpage{2439}
(\byear{2008})
\end{barticle}
\endbibitem

\bibitem[\protect\citeauthoryear{Kandasamy et~al.}{2018}]{kandasamy:2018}
\begin{bchapter}
\bauthor{\bsnm{Kandasamy}, \binits{K.}},
\bauthor{\bsnm{Krishnamurthy}, \binits{A.}},
\bauthor{\bsnm{Schneider}, \binits{J.}},
\bauthor{\bsnm{Poczos}, \binits{B.}}:
\bctitle{Parallelised bayesian optimisation via thompson sampling}.
In: \beditor{\bsnm{Storkey}, \binits{A.}},
\beditor{\bsnm{Perez-Cruz}, \binits{F.}} (eds.)
\bbtitle{Proceedings of the Twenty-First International Conference on Artificial
  Intelligence and Statistics}.
\bsertitle{Proceedings of Machine Learning Research},
vol. \bseriesno{84},
pp. \bfpage{133}--\blpage{142}
(\byear{2018})
\end{bchapter}
\endbibitem

\bibitem[\protect\citeauthoryear{Santner et~al.}{2003}]{sant:will:notz:2003}
\begin{bbook}
\bauthor{\bsnm{Santner}, \binits{T.J.}},
\bauthor{\bsnm{Williams}, \binits{B.J.}},
\bauthor{\bsnm{Notz}, \binits{W.I.}}:
\bbtitle{The Design and Analysis of Computer Experiments}.
\bpublisher{Springer},
\blocation{New York, NY}
(\byear{2003})
\end{bbook}
\endbibitem

\bibitem[\protect\citeauthoryear{Surjanovic and Bingham}{2023}]{simulationlib}
\begin{botherref}
\oauthor{\bsnm{Surjanovic}, \binits{S.}},
\oauthor{\bsnm{Bingham}, \binits{D.}}:
Virtual Library of Simulation Experiments: Test Functions and Datasets.
Retrieved March 14, 2023, from \url{http://www.sfu.ca/\~{}ssurjano}
(2023)
\end{botherref}
\endbibitem

\bibitem[\protect\citeauthoryear{McKay et~al.}{1979}]{mcka:cono:beck:1979}
\begin{barticle}
\bauthor{\bsnm{McKay}, \binits{M.D.}},
\bauthor{\bsnm{Conover}, \binits{W.J.}},
\bauthor{\bsnm{Beckman}, \binits{R.J.}}:
\batitle{A comparison of three methods for selecting values of input variables
  in the analysis of output from a computer code}.
\bjtitle{Technometrics}
\bvolume{21},
\bfpage{239}--\blpage{245}
(\byear{1979})
\end{barticle}
\endbibitem

\bibitem[\protect\citeauthoryear{Kushner}{1964}]{kushner:1964}
\begin{barticle}
\bauthor{\bsnm{Kushner}, \binits{H.J.}}:
\batitle{A new method of locating the maximum of an arbitrary multipeak curve
  in the prescence of noise}.
\bjtitle{Journal of Basic Engineering}
\bvolume{86}(\bissue{1}),
\bfpage{97}--\blpage{106}
(\byear{1964})
\end{barticle}
\endbibitem

\end{thebibliography}

\end{document}